# Enhancing Morpho-Kinematic analysis for Plant Water Stress Classification through Leaf Movements

**Walter Polilli[1,2], Alessio Antonini[3], Cristiano Platani[2], Fabio Stagnari[1], Angelica Galieni[2]**

[1] Department of Bioscience and Agro-Food and Environmental Technology, University of Teramo, Via Renato Balzarini 1, 64100 Teramo, Italy

[2] Research Centre for Vegetable and Ornamental Crops, Council for Agricultural Research and Economics, CREA-OF, Via Salaria 1, 63077 Monsampolo del Tronto, Italy

[3] Knowledge Media Institute, The Open University, Milton Keynes, MK7 6AA, United Kingdom

**Abstract**

Precise irrigation management requires robust classification of plant water stress. We expanded a morpho-kinematic (MK) framework that derives canopy-movement features from RGB time-lapse imaging evaluating how methodological refinements affect robustness and fine discrimination across four irrigation treatments representing distinct stress histories.

The study tested both a biological (Agg) *versus* an isogonal (Unif) sectoring of the canopy image, within an additive scheme where to the baseline (i.e. flattened MK features, A0) were sequentially added non-linear descriptors (A1), irrigation-context variables (i.e. dry time, A2), and their interactions with baseline (A3). The multi-class problem was decomposed in biologically meaningful binary tasks, and the final classification confronted an adaptive – to the performance obtained in the out-of-fold predictions inside the leave-one-sample-out validation framework – linear opinion pooling (ALOP) ensemble, evaluated across its full parameter space, against hierarchical cascades (HCC).

In our combined dataset from two sequential *Lactuca sativa* L. experiments (144 sample-days) ALOP median outperformed HCC in every configuration, while non-linear and contextual enrichments (A1–A2) produced consistent improvements in terms of prediction stability, variability (for ALOP), and balanced accuracy (BA). The highest balanced accuracy (median BA = 0.96) was reached under Unif scheme in A3, yet the Agg configuration in A2 achieved the best compromise between accuracy (median BA = 0.91) and robustness.

Concluding, this study identifies methodological pathways that strengthen resilience and transferability of movement-based water-stress classification, establishing a solid foundation for generalizable, low-cost phenotyping.

## 1. Introduction

Optimizing irrigation is a critical challenge in modern agriculture, where water scarcity and climate variability increasingly threaten crop productivity (Kay et al., 2021). Traditional methods for assessing plant water status include destructive biomass measurements, soil moisture sensors, and plant-based techniques (e.g., leaf water potential, stomatal conductance) and provide valuable insights while facing practical limitations in terms of cost, labor intensity, temporal resolution, and invasiveness (Kacira et al., 2002; Jones 2004). Remote and proximal sensing technologies offer promising alternatives for high-throughput, non-destructive monitoring (Galieni et al., 2021), yet their deployment often requires expensive equipment or technical expertise that limits accessibility (Pierpaoli et al., 2013), particularly for small-scale growers or research institutions with constrained budgets.

Among physiological responses to water deficit, leaf movements represent a promising phenotypic proxy for plant water status. Both morphological adjustments (e.g., wilting, rolling) and nastic movements (e.g., paraheliotropism, nyctinasty) are modulated by turgor pressure changes and hydraulic signals, making them sensitive indicators of soil water availability (van Zanten et al., 2010; Kadioglu et al., 2012; Puglielli et al., 2017a; Puglielli et al., 2017b; Geldhof et al., 2021; Zeng et al., 2024). Early work demonstrated the potential of monitoring leaf angle dynamics for stress detection in crops such as soybean (Oosterhuis et al., 1985; Kacira et al., 2002), but practical implementation has long been hindered by the difficulty of capturing fine-scale temporal dynamics in field or greenhouse conditions.

Recent advances in sensing technologies have revitalized interest in movement-based phenotyping. Terrestrial laser scanning and LiDAR systems enable precise 3D reconstruction of canopy geometry and tracking of diurnal leaf angle patterns, revealing drought-induced paraheliotropic responses in species such as soybean and potato (Herrero-Huerta et al., 2018; Geldhof et al., 2021; Mulugeta Aneley et al., 2023; Chakhvashvili et al., 2023; Chakhvashvili et al., 2024; Köhl et al., 2023). Alternative approaches include millimeter wave radar for damped vibration analysis (Cardamis et al., 2025) and digital contact sensors for real-time movement quantification (Geldhof et al., 2021). While these methods achieve high spatial or temporal resolution, their hardware costs and technical complexity remain barriers to widespread adoption.

Computer vision approaches using RGB imaging offer a valid low-cost alternative. Recent studies have demonstrated the feasibility of informing leaf movement from time-lapse sequences for water stress discrimination, as for example in tomato (Koike et al., 2024) and in lettuce (Polilli et al., 2026), leveraging both classical image processing (Ramos-Giraldo et al., 2020; Rehman et al., 2020) and deep learning architectures (Koike et al., 2024; Kovár et al., 2024). Novel optical flow algorithms tailored to challenging imaging conditions, such as occluded leaves or foggy environment, further enhance robustness (Rao & Kashif, 2024) increasing expectations for future field level implementation. The accessibility of consumer-grade cameras and the interpretability of RGB-based features position this approach as particularly attractive for resource-limited settings and mechanistic studies linking visual phenotypes to physiological processes.

Building on this foundation, Polilli et al. (2026) formalized the morpho-kinematic (MK) features: a set of characteristics – extracted from time-lapse RGB imaging via optical flow analysis within a time-fixed observation window – encoding canopy movement and spatial distribution with a built-in relation to the age of the leaves via a hard geometrical partitioning. These MK features enabled a framework for water stress classification in lettuce (*Lactuca sativa* L.) that achieved promising discrimination (balanced accuracy, BA = 0.93) among four irrigation treatments (well-watered control, chronic stress, mild and severe acute stress) and demonstrating proof-of-concept for MK-based phenotyping. However, the framework exhibited several limitations that constrained its robustness and generalizability. Firstly, the dataset comprised a single experiment with limited sample size (n = 72). This restricted statistical power and forced the use of summary statistics (mean, standard deviation, max, and min) and linear trend descriptors (slope, intercept, coefficient of determination) to represent each MK within the observation window, potentially overlooking non-linear kinematic patterns. Secondly, the hard geometrical sectoring scheme (isogonal 60°) was sensitive to canopy asymmetries, compromising model transferability. Thirdly, it leveraged hierarchical classification architecture that, while perfect for interpretability, is susceptible to error propagation. Finally, the study was confined to a single species (lettuce), a controlled environment, and a late growth stage (36 to 41 days after transplanting) leaving open the questions about transferability and generalizability.

To specifically address the limitations affecting the framework's analytical robustness and stability, this study evaluates four methodological enhancements to the baseline morpho-kinematic pipeline: (i) biological aggregation of sectors by leaf developmental stage (Agg) *versus* original isogonal partitioning (Unif); (ii) non-linear temporal descriptors capturing kinematic accelerations within the observation window; (iii) irrigation context variables encoding time since previous watering event and their interactions with MK features; and (iv) adaptive linear opinion pooling (ALOP) for multi-class fusion *versus* original hierarchical classification cascades (HCC). We adopted a progressive additive design that traces performance trajectories across feature space enrichment levels while comparing two sectoring schemes (Agg vs Unif) and two ensemble architectures (HCC vs ALOP), prioritizing the identification of effective sectoring-feature-ensemble combinations.

## 2. Materials and Methods

While a schematic overview of the complete data-processing and modelling pipeline is provided in Figure 1; where the diagram identifies and clearly confines the methodological variants evaluated in this study along with the original steps inherited from Polilli et al. (2026). A more comprehensive description of the methodologies follows below.

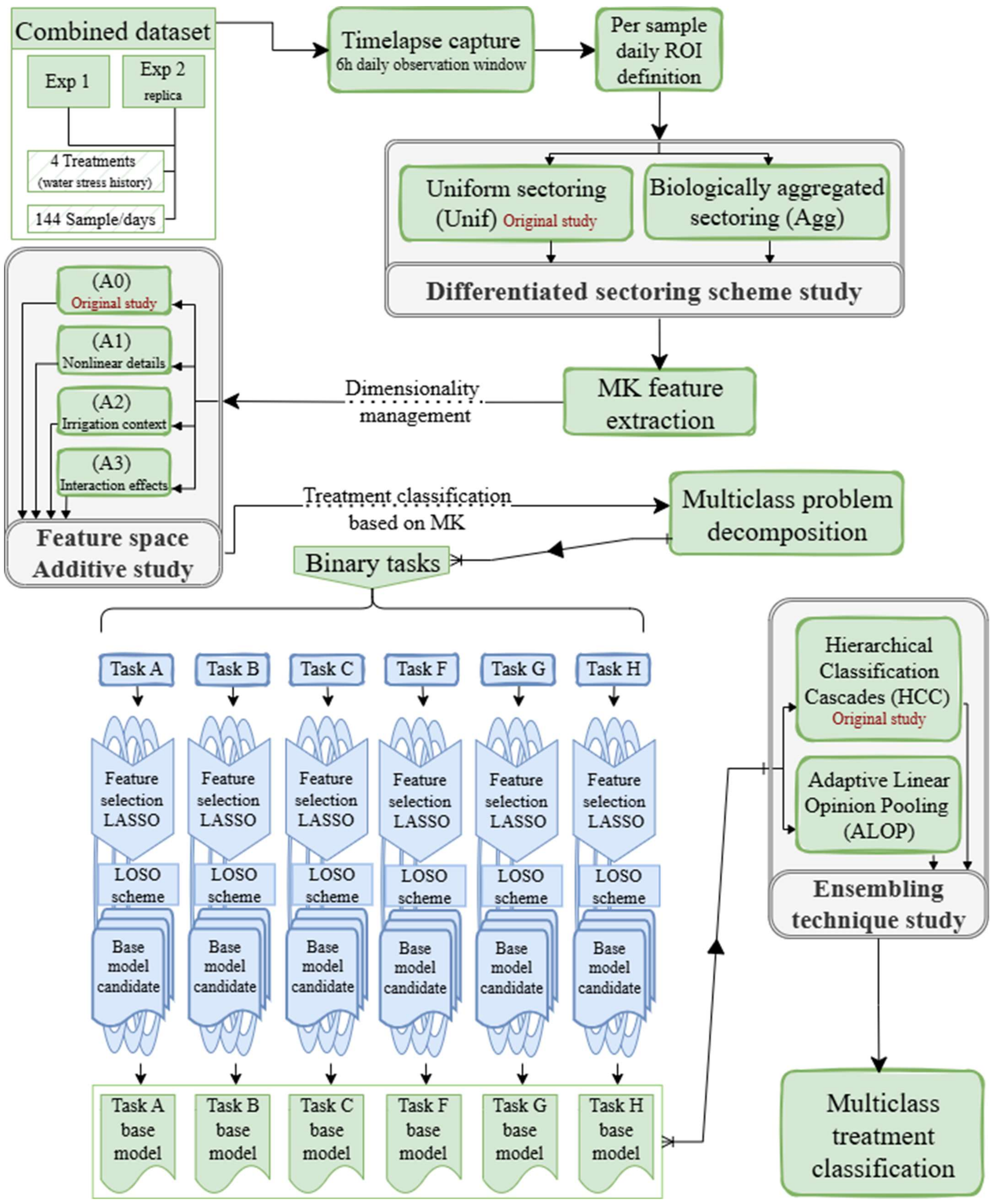

**Figure 1**: Overview of the study workflow and data-processing/modelling pipeline. Gray panels group the methodological comparisons (bold) and additions introduced in this work – steps from Polilli et al. (2026) are marked with a red reference. Blue blocks indicate where machine learning (ML) procedures were adopted. Green blocks, no ML involved.

## 2.1 Experimental Setup and Combined Dataset

This study builds on a combined dataset from two sequential experiments. The first is detailed in Polilli et al. (2026), while the second is a follow-up study designed to replicate its core methodology. In both experiments, lettuce plants (cv. Gretta Erre F3486.63) were grown in 3 L pots under controlled conditions (16-h photoperiod, 22 ± 1 °C). The substrate, fertilization schedule, and water stress treatments were identical, comprising a well-watered control (FC; 70% target Available Water Content, tAWC) and three water stress treatments (SC = chronic stress, SM = mild acute stress, SS = severe acute stress; all 25% tAWC) differing in the timing of stress imposition (starting at 15 or 28 Days After Transplanting, DAT) and irrigation frequency (every or every other day). An important modification in the follow-up experiment was a weekly one-day irrigation suspension occurred for all treatments throughout the entire growth cycle, contrasting with the single suspension event at 37 DAT in the first study. For a comprehensive description of soil properties, specific fertilizer applications, and treatment definitions, the reader is referred to Polilli et al. (2026).

We monitored 12 biological samples (3 replicates per 4 treatments) per experiment for 6 days during late growth stage (36-41 DAT), yielding a balanced dataset of 144 sample-days across four treatments. In both experiments, pictures were acquired using three USB camera modules (model HBV-1609 S2.0, manufactured by Thincol, China, mounting a OV2643 1/4" CMOS sensor, Automatic Exposure Control – AEC, a 120° FOV, and an F/2.0 fixed-focus lens). To harmonize datasets, we applied z-score standardization within each outer training fold only (parameters estimated on training data and applied to the held-out replicate); the common cross-validation folding scheme is described below.

## 2.2 Image Processing

The time lapse building, image processing, and feature extraction pipeline was mostly replicated from Polilli et al. (2026). The observation window (6 hours) starts at the daily pot weighing event (that may or may not result in sample irrigation) and during such a window 25 JPEG pictures were captured, one every 15 minutes (see Supplementary Figure S1). Prior to further analysis, all images from both experiments underwent a camera-specific lens distortion correction. This transformation was calibrated empirically by adjusting the primary radial distortion coefficients (barrel and pincushion) until known rectilinear elements within the scene visually aligned with superimposed reference lines (Fernandes et al., 1997).

A circular daily Region of Interest (ROI) was established for each sample/day by centering it on the daily average centroid and defining its radius (Rk) as the maximum Euclidean distance from the centroid observed that day. Specifically, the first step was to segment images into a binary canopy mask using thresholding on the Excess Green index (Arroyo et al., 2016) followed by morphological opening (5x5 pixels). Then, let $n \in \{1, \dots, 25\}$ be the time index of the captured frames, and $I_{\text{bin}}(x, y, n) \in \{0,1\}$ be the binary canopy mask at time $n$: for each frame, the largest connected component is identified and used to compute the canopy centroid:

$$\mathbf{c}(n) = (c_x(n), c_y(n)).$$

For a given sample $k$and day $d$, the daily mean centroid is:

$$\bar{\mathbf{c}}_{k,d} = \frac{1}{T}\sum_{n=1}^{T} \mathbf{c}(n),$$

where $T = 25$ is the number of frames in the observation window.

For each time $n$, the maximum Euclidean distance of canopy pixels from the daily mean centroid was defined as follows:

$$r(n) = \max_{(x,y): I_{\text{bin}}(x,y,n)=1} \sqrt{(x - \bar{\mathbf{c}}_x)^2 + (y - \bar{\mathbf{c}}_y)^2}$$

The daily ROI radius is:

$$R_{k,d} = \max_{T} r(n).$$

The ROI is the circular mask centered at $\bar{\mathbf{c}}_{k,d}$with radius $R_{k,d}$.

This approach provided a consistent spatial reference for movement analysis operating throughout optical flow (Farneback algorithm). A dynamic visualization of this process is provided by Supplementary Video S2 and its caption (see “Supplementary Materials” file).

### 2.3 Base Morpho-Kinematic (MK) Features Extraction

MK analysis involves the sectorization of the ROI in order to independently monitor the dynamics of tissues that are expected to exhibit different behaviors in response to different water stress histories. We compared two sectoring approaches while progressively enriching the baseline time-series flattening features (Polilli et al., 2026). Sectoring approaches (Figure 2):

- Uniform Sectoring (Unif): The original isogonal (60°) sectors, named clockwise S1 to S6.
- Biologically aggregated Sectoring (Agg): sectors S1, S2, and S3 became a single continous 180° old leaves sector (SOL), sectors S4 and S6 became a discontinuous 120° sector capturing young and developed leaves (SYD), while sector S5 geometry, capturing central rosette, was unmutated and renamed SCR.

With the use of optical flow, for each sector and each timestep $t$ (the interval between $n = t$ and $n = t + 1$) in the observation window, the six base MK features were computed (Table 1). Figure 3 deploys a visual aid for MK extraction in a single timestep, while detailed formulas and algorithm parameters, are reported in Supplementary Table S4.

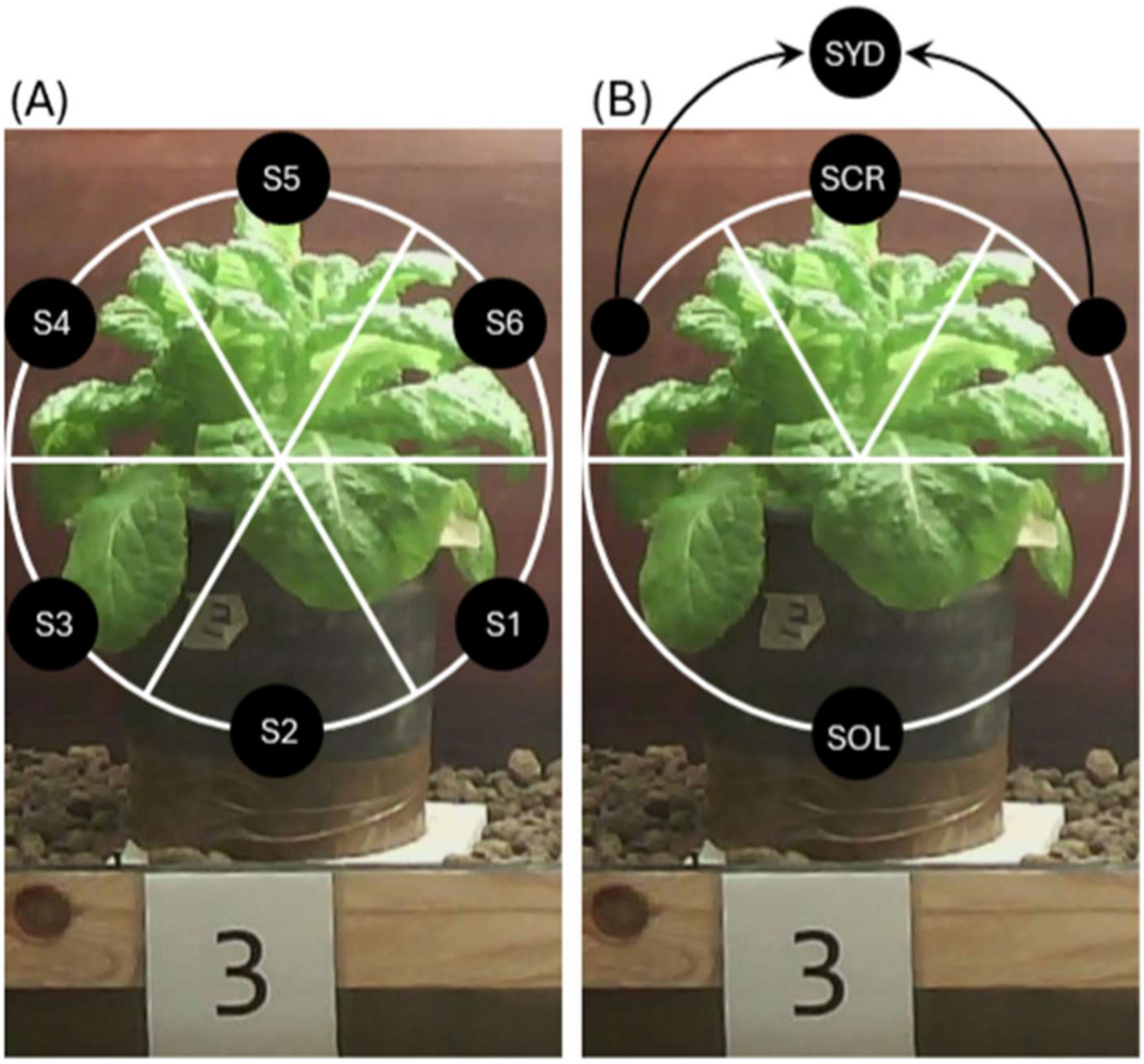

**Figure 2:** Visualization of the sectoring approaches, panel A uniform isogonal (Unif) as defined in Polilli et al. (2026); panel B biological aggregated (Agg). Sector young and developed leaves (SYD) includes both the chunks connected by black dots and arrows.

**Table 1:** Base morpho-kinematic (MK) features computed per sector within the daily ROI. Formal definitions with equations as well as thresholds and normalization parameters are provided in Supplementary Table S4. The Purpose column reflects the interpretative key given the assumption that sectoring scheme encodes for different leaf age classes.

| Extended Name | Abbreviation | Operational definition | Purpose |
|---|---|---|---|
| Canopy Density | D | Fraction of ROI sector area classified as canopy in a frame | Spatial distribution, morphological canopy architecture |
| Movement Threshold | Mth | Fraction of ROI sector pixels exceeding motion threshold between consecutive frames (optical flow magnitude) | Broad extent of active motion associated with leaf age |
| Mean Vertical Movement | VM | Mean signed vertical displacement of moving pixels between consecutive frames (normalized) | Directional vertical component of motion, associated with leaf age |
| Mean Horizontal Movement | HM | Mean signed horizontal displacement of moving pixels between consecutive frames (normalized) | Directional horizontal component of motion, associated with leaf age |
| Mean Movement Magnitude | MMag | Mean optical-flow magnitude of moving pixels between consecutive frames (normalized) | Overall motion intensity independent of direction, associated with leaf age |
| Movement Magnitude Coefficient of Variation | MagCV | Coefficient of variation of optical-flow magnitudes over moving pixels (normalized) | Heterogeneity/dispersion of motion within the same leaf age class – *i.e.* few moving leaves |

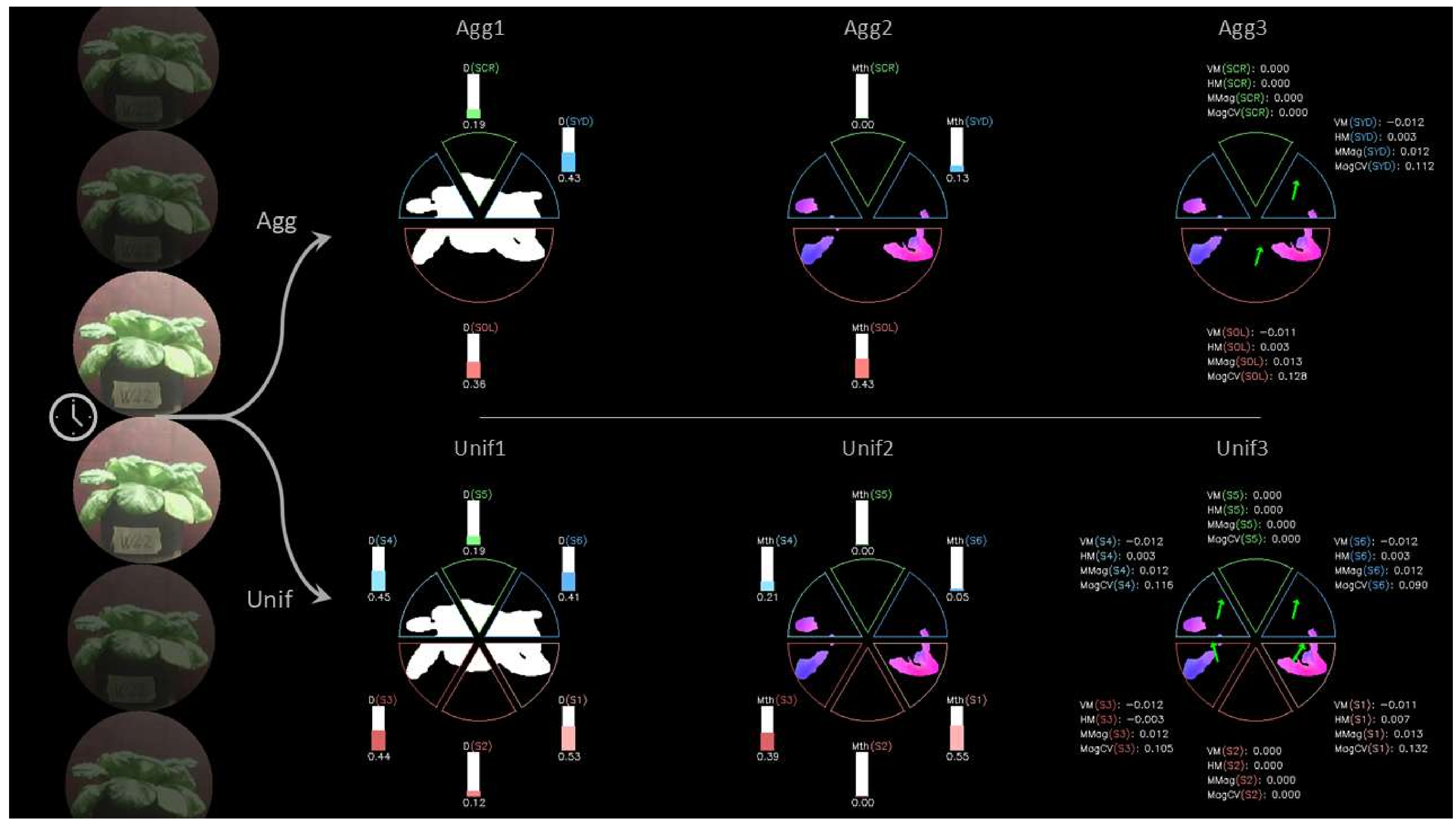

**Figure 3**: Workflow for Morpho-Kinematic (MK) computation. Example for one sample-day (single observation window) and one timestep *t* following both Uniform (Unif) and Biological aggregated Sectoring (Agg). At first the canopy mask is involved for calculating D; thereafter, the thresholded flow depicts Mth. In the final step a movement vector, averaged throughout the sector, finally appears to visualize VM, HM, and MMag. MagCV can be inferred by differences in the saturation of the flow color map (the flow color map derived from HSV color space with H defining direction and S magnitude). For MK acronyms refer to Table 1.

## 2.4 Additive Feature Engineering Study

In the present study, the dataset size - although did not allow the direct use of raw MK features - enabled the evaluation of more advanced flattening approaches with respect to Polilli et al. (2026). Accordingly, we explored an expansion of the feature set through additive modeling strategies. In particular, for both sectorial schemes (i.e., Unif and Agg), we evaluated four nested feature sets:

- Baseline (A0): The flattened summary statistics of the base MK features (i.e., mean, std, min, max, slope, intercept, rsq).
- Nonlinear details (A1): A0 features plus the slope of the linear regression fit on the first and third tertiles of each daily MK time-series.
- Irrigation Context (A2): A1 features plus variables encoding the time delta ($\Delta t$) since the last previous irrigation event ($\Delta t$, $\Delta t^2$, and $\log(\Delta t)$). Where $\Delta t$ is the elapsed time – normalized 0 to 1 - from the last irrigation to the daily pot weighing; not reset on same-day irrigation.
- Interaction Effects (A3): A2 features plus the most significant interaction terms between irrigation context variables and the MK features from A1.

Visual examples for the additive feature engineering levels A0-A2 are provided in Supplementary Figure S5.

## 2.5 Problem Decomposition and Task Definition

Following the approach in Polilli et al. (2026), the 4-class classification problem (FC, SC, SM, SS) was decomposed into a series of simpler binary classification sub-problems, hereafter referred to as "tasks". The selection of these tasks was guided by a dual rationale. The primary driver remained their physiological interpretability, with each task designed to test a specific biological hypothesis (e.g., distinguishing acute vs. chronic stress). The secondary driver was to assemble a set of base classifiers suitable for two purposes: re-evaluating the most effective Hierarchical Classification Cascades (HCC) from our previous work and constructing a novel Adaptive Linear Opinion Pooling (ALOP) ensemble based on an Error-Correcting Output Codes (ECOC) framework (Dietterich & Bakiri, 1994) The six binary tasks defined for this study are detailed in Table 2.

**Table 2:** treatments description with watering schedule followed by binary classifiers (task) description.

| Treatment | Irrigation schedule | | Representing |
|---|---|---|---|
| FC | 1 to 42 DAT - 75% tAWC daily | | Well-watered |
| SC | 1 to 15 DAT - 75% tAWC daily<br>16 to 28 DAT - 25% tAWC daily<br>29 to 42 DAT - 25% tAWC every other day | | Chronic water stress |
| SS | 1 to 28 DAT - 75% tAWC daily<br>29 to 42 DAT - 25% tAWC every other day | | Acute severe stress |
| SM | 1 to 28 DAT - 75% tAWC daily<br>29 to 42 DAT - 25% tAWC daily | | Acute mild stress |
| **Task** | **Positive** | **Negative** | **Representing** |
| A | FC | {SC, SS, SM} | Well-watered vs any stress |
| B | SC | {SS, SM} | Chronic vs acute stresses |
| C | SM | SS | Mild vs severe stress |
| F | FC | SC | Well-watered vs chronic stress |
| G | SC | {FC, SS, SM} | Chonic stress vs well-watered history |
| H | {FC, SC} | {SS, SM} | Adapted vs responding |

### 2.5 Feature Selection and Base Model Training

For each task, we selected predictive features via frequency-based (50% occurrence) Least Absolute Shrinkage and Selection Operator (LASSO) within a logistic regression in a Leave-One-Sample-Out (LOSO) framework. Each outer fold excluded one biological replicate with all its daily observations (36–41 DAT), ensuring no observation from the same plant appears simultaneously in training and test sets. Using the same LOSO loop, the selected feature sets were utilized to train and evaluate eight different types of classifiers for each task: Logistic Regression (LogReg), K-Nearest Neighbours (KNN), Support Vector Classifier with Radial Basis Function kernel (SVC-RBF), Support Vector Classifier with Linear kernel (SVC-Linear), Random Forest (RF), Extreme Gradient Boosting (XGB), Ligth Gradient Boosting Machine (LGBM), and Linear discriminant analysis (LDA). Model hyperparameters were tuned via a 3-fold cross-validation nested in the LOSO loop, optimizing for BA. This process yielded a set of validated binary classifiers, which served as the foundational components for the ensemble architectures described below.

### 2.6 Comparative Analysis of Ensemble Architectures

Operating on the same set of binary base classifiers and without training an additional supervised meta-learner, we compared two multi-class ensemble frameworks as follows. For each task, we first generated out-of-fold (OOF) probability estimates from all trained architectures across LOSO folds. Among all combinations of additive levels and sectoring approaches, we then selected, for each task, the single architecture achieving the best BA. The corresponding OOF probability estimates were then used as inputs for both ensemble architectures described below.

- Hierarchical Classification Cascade (HCC): The sequential classification architecture from Polilli et al. (2026) was re-evaluated. This framework provides a final classification through a rule-based decision tree, offering maximum interpretability. We focused on the three previously identified effective cascade structures: A→B→C, G→A→C, and H→F/C, with the letters referring to tasks presented in Table 2.
- Adaptive Linear Opinion Pooling (ALOP): Methods for combining predictions from multiple classifiers range from stacked generalization (Wolpert, 1992) to weighted voting and averaging schemes (Kuncheva & Rodríguez, 2014; Large et al., 2019). Our ALOP framework belongs to the latter family, combining outputs via performance-based weighting. This approach calculates a final four-class probability distribution by aggregating the predicted probabilities from the set of binary base classifiers.

The ALOP core mechanism operates as follows: (i) the contribution of each base classifier to a class's score is determined by its task definition. If a given class (e.g., FC) is the positive target for a task, the model's output probability for the positive target ($p$) contributes to that class's score; if the class belongs to the negative set, the complementary ($1 - p$) is used instead. Two indicator matrices, $\delta P$ and $\delta N$, encode these relations between tasks and classes, determining whether each classifier contributes positively, negatively, or not at all to a given class. (ii) Each class score $S_j$ is then computed as a weighted linear combination of the relevant classifier outputs with the following equation:

$$S_j = \sum_i \; w_i(\delta P_{ij} p_i + \delta N_{ij}(1 - p_i))$$

Where $w$ represents the adaptive weights, $i$ iterates over base models and $j$ over classes. This process yields a vector of four accumulated class scores, each obtained through a weighted averaging scheme belonging to the Hölder family of fusion rules (arithmetic case; Koliander et al., 2022). Finally, a linear normalization is applied to normalize the score vector into a valid probability distribution where all class probabilities sum to one (Figure 4).

A systematic grid search was conducted to evaluate the ALOP framework. The objective was not to identify a single optimal configuration – a process prone to overfitting and winner's bias – but rather to assess the architecture's performance robustness across a range of structural parameters.

We explored: (i) parameters for calibrating probability estimates, that were either used directly (identity) or rectified using Platt scaling, isotonic regression, and beta calibration. These techniques are for correcting systematic distortions in predicted probabilities: Platt scaling addresses sigmoid-shaped distortions, while beta calibration handles both under and overconfident predictions (Niculescu-Mizil & Caruana, 2005; Kull et al., 2017). (ii) Weighting schemes based on the LOSO OOF performance of the base models, within a framework very close to the cross-validation accuracy weighted probabilistic ensemble (CAWPE) proposed by Large et al. (2019) – differing for the $\alpha = 1$ in our case and $\alpha = 4$ in CAWPE – where we confronted uniform weights, referred as "none" ($\alpha = 0$), with a larger pool of performance-based and combined metrics (i.e., none, BA, BA*(1-ECE), 1-brier, BA*(1-brier), BA*AUC). A $\gamma$ parameter $\in \{1.0, 2.0\}$ was added to the test to modulate the influence of the weights. (iii) Two pruning strategies to exclude low-performing classifiers from the pool: TopK $\in$ {"all", 3} defining the number of admitted base models based on their BA ranking, and Task Exclusion $\in \{0.00, 0.80\}$ as a BA threshold for admission. And (iv) the probabilities estimates were combined using either raw or squared arithmetic means.

| | i = Task A | i = Task B | i = Task C | i = Task F | i = Task G | i = Task H |
|---|---|---|---|---|---|---|
| j = FC | $\delta P_{ij} = 1$<br>$\delta N_{ij} = 0$ | $\delta P_{ij} = 0$<br>$\delta N_{ij} = 0$ | $\delta P_{ij} = 0$<br>$\delta N_{ij} = 0$ | $\delta P_{ij} = 1$<br>$\delta N_{ij} = 0$ | $\delta P_{ij} = 0$<br>$\delta N_{ij} = 1$ | $\delta P_{ij} = 1$<br>$\delta N_{ij} = 0$ |
| j = SC | $\delta P_{ij} = 0$<br>$\delta N_{ij} = 1$ | $\delta P_{ij} = 1$<br>$\delta N_{ij} = 0$ | $\delta P_{ij} = 0$<br>$\delta N_{ij} = 0$ | $\delta P_{ij} = 0$<br>$\delta N_{ij} = 1$ | $\delta P_{ij} = 1$<br>$\delta N_{ij} = 0$ | $\delta P_{ij} = 1$<br>$\delta N_{ij} = 0$ |
| j = SS | $\delta P_{ij} = 0$<br>$\delta N_{ij} = 1$ | $\delta P_{ij} = 0$<br>$\delta N_{ij} = 1$ | $\delta P_{ij} = 0$<br>$\delta N_{ij} = 1$ | $\delta P_{ij} = 0$<br>$\delta N_{ij} = 0$ | $\delta P_{ij} = 0$<br>$\delta N_{ij} = 1$ | $\delta P_{ij} = 0$<br>$\delta N_{ij} = 1$ |
| j = SM | $\delta P_{ij} = 0$<br>$\delta N_{ij} = 1$ | $\delta P_{ij} = 0$<br>$\delta N_{ij} = 1$ | $\delta P_{ij} = 1$<br>$\delta N_{ij} = 0$ | $\delta P_{ij} = 0$<br>$\delta N_{ij} = 0$ | $\delta P_{ij} = 0$<br>$\delta N_{ij} = 1$ | $\delta P_{ij} = 0$<br>$\delta N_{ij} = 1$ |

$$S_j = \sum\nolimits_i w_i \left(\delta P_{ij} p_i + \delta N_{ij}(1 - p_i)\right)$$

$$\begin{bmatrix} S_{FC} \\ S_{SC} \\ S_{SS} \\ S_{SM} \end{bmatrix} \xrightarrow[\text{normalization}]{\pi_j = \frac{S_j}{\sum_j S_j}} \begin{bmatrix} \pi_{FC} \\ \pi_{SC} \\ \pi_{SS} \\ \pi_{SM} \end{bmatrix}$$

**Figure 4:** Adaptive Linear Opinion Pooling (ALOP) mechanism and class-score aggregation process. The top panel shows the task–class encoding matrices δP and δN, indicating whether each binary classifier contributes positively (δP = 1), negatively (δN = 1), or not at all (0) to each target class. The bottom panel illustrates the computation pipeline: weighted accumulation of classifier outputs into class scores (Sj). Other equation parameters are i iterating over base models and j over classes; w = adaptive weights, p = estimated probability for the positive class. A final normalization is applied to obtain the four-class probability estimates (πj).

## 3. Results

### 3.1 Efficacy of Stress Treatments and Robustness of the Baseline (A0) Feature Space

In both experiments water stress was imposed differentially, at 15 DAT for treatment SC and at 28 DAT for treatments SS and SM. To let the plants fully develop a differentiated stress history, the observation window was set after an acclimation period between 36 and 41 DAT, and to assess the effectiveness of treatments this window was preceded and followed by two destructive samplings at 35 and 42 DAT. We analyzed lettuce fresh weight via two-way ANOVA (Type III, experiment as blocking factor) with heteroskedasticity-robust HC3 inference due to borderline Breusch-Pagan test results. This analysis resulted in a significant effect of treatment (*p*-value <0.001), whereas experiment, sampling DAT, and treatment×sampling were not (*p*-values 0.0510, 0.9918, 0.8504, respectively).

The analysis of the baseline feature space (A0), obtained by summarizing the daily time-series with descriptive and linear regression statistics, allowed assessing the robustness of the core MK signatures identified in previous work (Polilli et al., 2026), and most observations were confirmed. (i) Feature selection was dominated by parameters derived from linear regression – especially linear trend stability (rsq) – over descriptive statistics. (ii) The dynamics of Canopy Density (D) in older leaf sectors (SOL in the Aggregated approach) remained a universally critical predictor, with a preeminent co-occurrence of magnitude descriptive flattening in Tasks A and F. (iii) Complex and broad tasks focusing on stress discrimination (e.g., Task B, G, H) consistently selected larger and more diverse feature sets. (iv) Tasks requiring finer discrimination (e.g., Task C: SM vs. SS) relied more heavily on kinematic features (e.g., MMag, HM, VM) over morphological ones (e.g., D). Conversely, two behaviors did not stand to the enlarged dataset: (i) central rosette (S5 or SCR) was not systematically over-represented in Tasks B, F, and G, as well as (ii) movement threshold preference for descriptive statistics was not confirmed. For a full overview of the selected features please see Supplementary Table S6.

### 3.2 Enhancing the Feature Space: Comparing Sectoring Approaches and Additive Levels

Building upon the baseline A0 space, we evaluated two methodological enhancements: a physiologically driven sectoring approach (Agg) compared to the original isogonal one (Unif) (see Figure 1), and the incremental addition of feature levels (A1-A3).

The Agg approach proved to be more parsimonious. While at the A0 level the number of selected features was comparable between the two approaches, at higher additive levels (A2-A3) Agg tended to select a smaller, more concise feature set (Figure 5). Adding the first enrichment level (A1) revealed the importance of non-linear dynamics within the daily observation window. In the baseline A0 level, rsq was a dominant feature, making up 24.0% and 29.9% of the selected features for the Agg and Unif approaches, respectively.

With the addition of slope parameters calculated on the external temporal thirds of the window (slope_t1 and slope_t3) in A1, the analysis gained better insights on MK accelerations or decelerations. The importance of these new indicators is clear: in the Agg approach, the tertiles' slopes became preeminent, accounting for 25.4% of selections and surpassing rsq, whose share dropped to 18.9%. In the Unif approach, a substantial balance was observed between the linear trend stability (25.2%) and the new slope_t1 or t3 (24.8%), together accounting for exactly half of the selected flattening.

This preference was also task specific. In the Agg models, tertiles' slopes were favored over rsq in all tasks that discriminated between different stress types. The only exceptions were Tasks A and F, those which identify the well-watered (FC) treatment. By contrast Task G – chronic stress (SC) vs all other treatments, showed the greatest use of these non-linear features in both sectoring approaches and in the following addition levels A2 and A3 (shares varying from 24.0% to 34.0% of selected features) – see Supplementary Table S6.

At level A2, we integrated the time elapsed since the previous irrigation ($\Delta t$) along with its squared and logarithmic transformations. It is important to note that treatments differed in irrigation frequency, with two groups sharing identical schedules ({FC, SM} daily; {SS, SC} every other day), that means that $\Delta t$ carries a schedule-aligned signal enabling partial separation of these groups (AUC = 0.69 ± 0.08; see Figure 6 and Supplementary Table S7). To evaluate the nature of $\Delta t$'s signal, we categorized the tasks based on their classes' alignment with the irrigation schedule in schedule-aligned (Task F and C), partially aligned (Task A, B, and G), and schedule-orthogonal (Task H).

A signal derived purely from scheduling would predictably yield the highest performance gains in the aligned tasks, and none in the orthogonal ones. However, three observations suggest $\Delta t$ provides additional information beyond scheduling: (i) feature selection retained a $\Delta t$-related feature in all tasks regardless of how task classes aligned with irrigation schedules (Supplementary Table S6); (ii) performance gains from A1 to A2 (Figure 7) were heterogeneous across tasks, with substantial improvements observed even where schedule-based separation was not expected; (iii) different tasks favored different $\Delta t$ transformations [raw $\Delta t$ vs $(\Delta t)^2$], suggesting context-dependent informativeness. A detailed analysis supporting the interpretation of $\Delta t$ as carrying both schedule-aligned and context-defining components is provided in Appendix A.

In A3, we explicitly added interaction terms formed by multiplying the irrigation context variables ($\Delta t$, $\Delta t^2$, log $\Delta t$) with the MK summaries, to test whether direct $\Delta t \times$MK effects would improve performance beyond A2. Figure 5 reports, for each task and level, the cardinality of the robust feature set and its composition by source level (A0, A1, A2, A3).

Under the Agg scheme, three facts are observed. First, the size of selected features sets decreases from A2 to A3 across all tasks. Second, mostly no A3-native interaction terms are retained by the robustness criterion (i.e., occurrence threshold across LOSO folds) in the A3 feature sets (only one is retained in Task F). Third, A3 shows a sparse and reduced BA advantage relative to previous levels – see Table 3.

Conversely, under the Unif scheme, A3 occasionally enlarges the robust sets and can introduce interaction terms in some tasks; however, these occurrences are not systematic across tasks and do not yield a consistent advantage over A2 (Table 3).

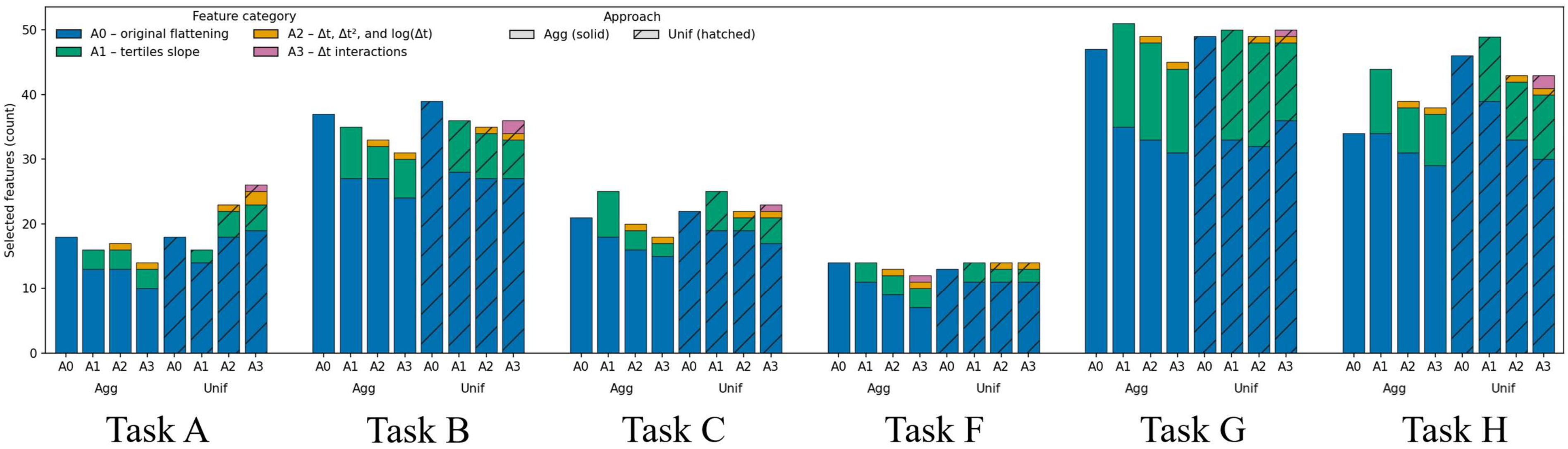

**Figure 5:** Stacked bar chart showing, for each task (refer to Table 1 for task definition), the number of selected features at each additive level (A0–A3) under two sectoring approaches (Agg and Unif). Colors encode feature category: A0 = original flattening; A1 = tertiles slope; A2 = Δt transform; A3 = Δt transform interactions. Pattern encodes approach (Agg = solid, Unif = hatched).

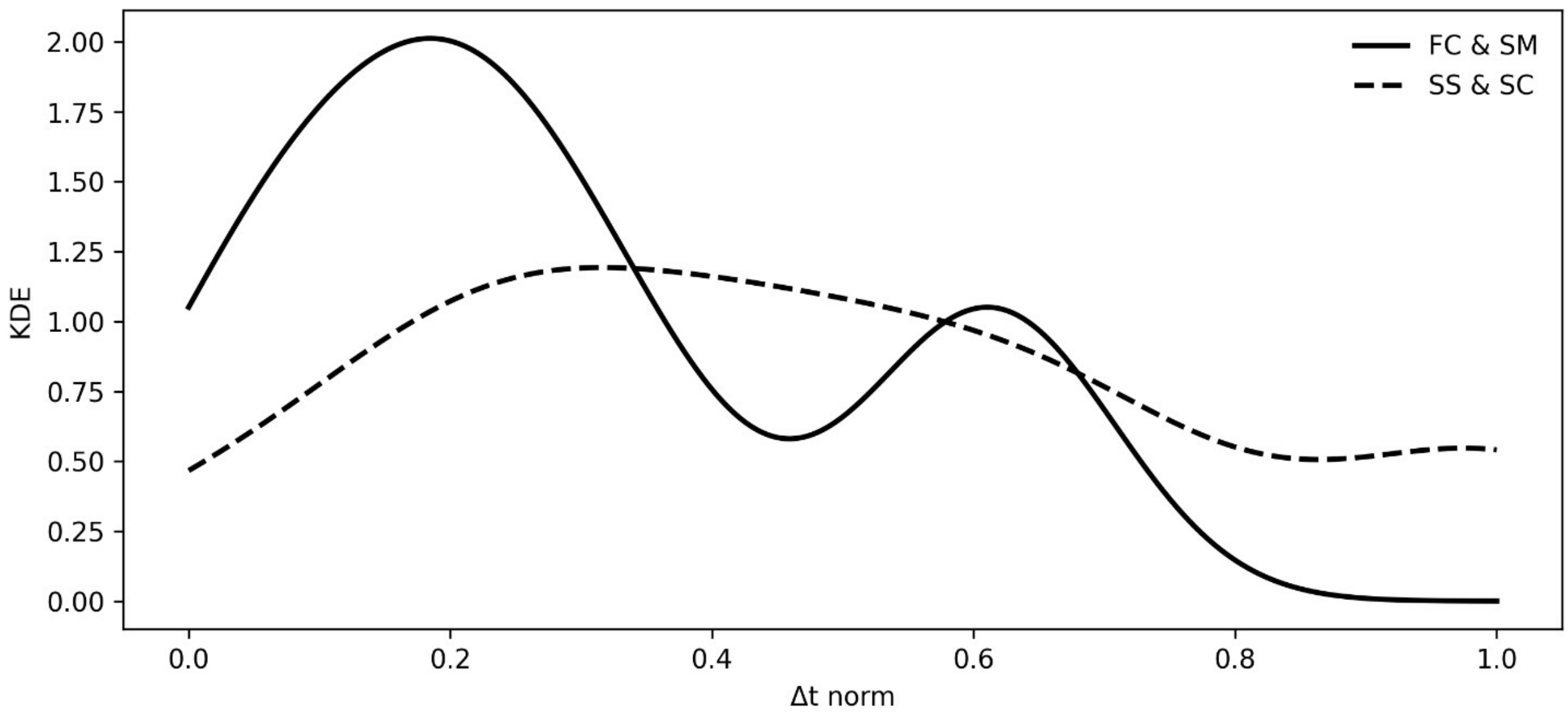

**Figure 6:** Kernel smoothing for probability density estimation (KDE) of normalized Δt (Δt norm) values for the two treatments groups sharing the same irrigation schedule ({FC, SM} daily; {SS, SC} every other day).

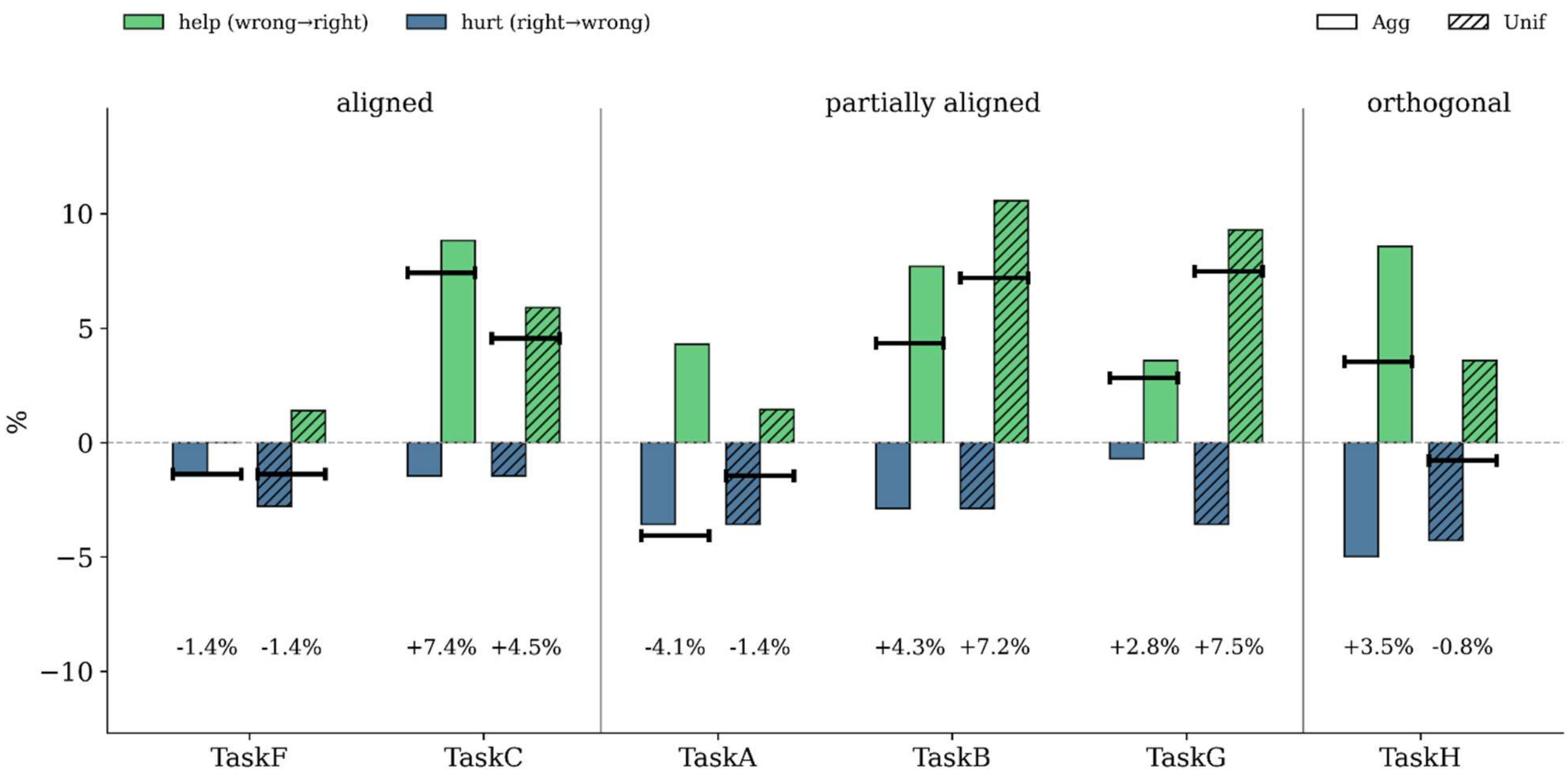

**Figure 7:** Barcharts show sample classification switches from additive level A1 to A2 separating "help" (wrong → right; green) and "hurt" (right → wrong) counts (percentage of total samples) for all tasks. Pattern encodes approach (Agg = solid, Unif = hatched). Horizontal black segments with percentage labels on the lower side of the plot represent balanced accuracy gain in A2 vs A1. Panels divide tasks according to their class alignment with irrigation schedule (as described in Results 3.2).

**Table 3:** Per each sectoring scheme, Task and additive level table shows architecture; Balanced Accuracy; Area Under the receiver operator Curve (both calculated on out-of-fold predictions) of the base-model chosen for subsequent ensemble techniques evaluation.

| Sectoring scheme | Task | A0 | A1 | A2 | A3 |
|---|---|---|---|---|---|
| Agg | TaskA | SVC_RBF; 0.88; 0.93 | KNN; 0.95; 0.95 | SVC_Linear; 0.91; 0.98 | KNN; 0.95; 0.96 |
| | TaskB | LDA; 0.80; 0.86 | LDA; 0.83; 0.90 | LDA; 0.87; 0.94 | LDA; 0.87; 0.93 |
| | TaskC | LogReg; 0.84; 0.90 | LogReg; 0.88; 0.93 | SVC_RBF; 0.96; 0.99 | SVC_RBF; 0.91; 0.95 |
| | TaskF | XGB; 0.93; 0.96 | LogReg; 0.94; 0.95 | LogReg; 0.93; 0.95 | LogReg; 0.96; 0.98 |
| | TaskG | LogReg; 0.68; 0.78 | LDA; 0.80; 0.88 | LDA; 0.83; 0.89 | LogReg; 0.84; 0.86 |
| | TaskH | LogReg; 0.86; 0.93 | SVC_Linear; 0.84; 0.91 | LDA; 0.88; 0.95 | LogReg; 0.85; 0.95 |
| Unif | TaskA | KNN; 0.93; 0.97 | LDA; 0.95; 0.98 | LDA; 0.94; 0.98 | LDA; 0.94; 0.98 |
| | TaskB | LogReg; 0.78; 0.85 | LogReg; 0.78; 0.84 | SVC_Linear; 0.82; 0.88 | LogReg; 0.83; 0.89 |
| | TaskC | LogReg; 0.83; 0.85 | LogReg; 0.84; 0.89 | LogReg; 0.91; 0.96 | LogReg; 0.91; 0.97 |
| | TaskF | LGBM; 0.95; 0.99 | LogReg; 0.97; 0.98 | LogReg; 0.97; 0.97 | LogReg; 0.96; 0.98 |
| | TaskG | LogReg; 0.73; 0.80 | LogReg; 0.74; 0.82 | LogReg; 0.78; 0.86 | LogReg; 0.73; 0.82 |
| | TaskH | LDA; 0.87; 0.93 | SVC_RBF; 0.84; 0.91 | LDA; 0.86; 0.94 | LogReg; 0.86; 0.93 |

### 3.3 Impact of the Enhanced Feature Space on Ensemble Classifier Performance

The four feature spaces were used in a multi-class classification context, comparing the performance of the previously developed hierarchical classification cascade (HCC) with a new adaptive linear opinion pooling (ALOP) approach.

Figure 8 indicated that ALOP outperformed HCC across all combinations of sectoring schemes and additive levels. The median BA advantage of ALOP over the best-performing cascade path (among ABC, GAC, HFC) was consistent under both sectoring schemes and ranged from 1 to 10 percentage points (p.p.), with the largest gains observed for Agg (A1 10 p.p.).

Examining BA trajectories across additive levels, the Unif approach exhibited a steady increase in ALOP median BA from 0.86 (A0) to a peak of 0.96 (A3), accompanied by a progressive tightening of the interquartile range, indicating robust behavior across pooling configurations. The Agg variant followed a similar yet less pronounced trend, improving from 0.80 (A0) to 0.91 (A2), while A3 yielded a decrease (0.89), mirroring the limited marginal contribution of A3-level interaction terms observed in Section 3.2.

The cascades showed consistent dynamics: under Unif, the best sequence reached 0.906 (A3), whereas under Agg it plateaued at 0.843 (A2) and declined at A3 (0.815), always below the corresponding ALOP medians. Table 4 provides insights from the ALOP Grid Search enlightening the mechanics behind these trends. (i) At higher additive levels, uncalibrated (identity) base probabilities consistently delivered the best results — in Unif A3 and Agg A1–A3 — implying that the enriched feature space and the underlying models already generate well-structured probabilistic gradients. (ii) Performance-adaptive weighting schemes, such as BA×(1 − Brier) in Agg A2 and BA×AUC in Unif A3, often stabilized or enhanced BA at advanced levels, especially when combined with Top-3 model selection. (iii) The task-exclusion criterion proved effective only in Agg A0 as in other combinations no base models performed worse than the threshold value.

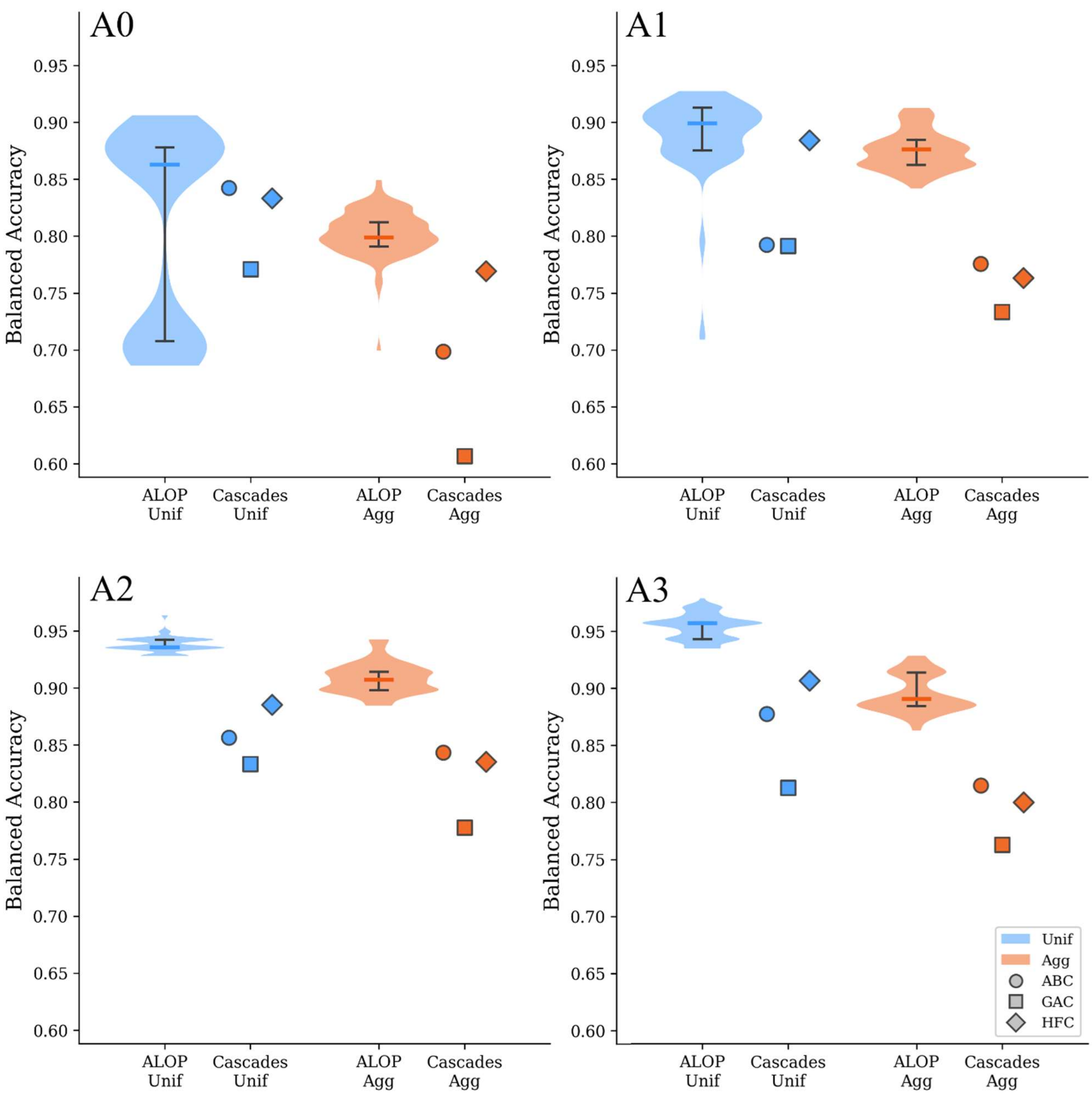

**Figure 8**: Comparison of ALOP (violin plots) distributions and HCC (individual markers) performance (balanced accuracy) across additive levels (A0-A3) for two sectoring strategies: Unif and Agg. Violin plots display quartile markers: horizontal black lines indicate Q1 and Q3 connected by a vertical segment, while the median is shown as a colored horizontal line. The three cascade architectures are represented by distinct geometric shapes (circle: A→B→C, square: G→A→C, diamond: H→F/C). Colors distinguish the two sectoring strategies.

**Table 4:** Summary of parameters main effects and the top two-way interaction per approach × addition level on adaptive linear opinion pooling (ALOP) prediction performance (refer to section 2.3 for addition levels and sectoring scheme, and to section 2.6 for ALOP parameters). Each cell reports on separate lines: the best value for the parameter, z-score (Δ vs 2nd best), and p-value. Main effects are shown only if $p \le 0.05$ and $\Delta z \ge 0.1$ ("--" = not achieved). The Top interaction column displays the best interaction setting (e.g., 'WeightScheme=uniform & TopK=3') only if $p \le 0.05$, partial $\eta^2 \ge 0.05$, and $\Delta z \ge 0.1$. Balanced Accuracy is standardized within each approach × addition level.

| Sectoring scheme × Addition level | Calibration | Weighting scheme | Gamma | Task Exclusion | TopK | Top interaction |
|---|---|---|---|---|---|---|
| Unif × A0 | -- | **none**<br>z: +0.91 (Δ +0.89)<br>p: 2.72E-12 | -- | -- | **all**<br>z: +0.90 (Δ +1.98)<br>p: 0.00E+00 | -- |
| Unif × A1 | **beta**<br>z: +0.78 (Δ +0.43)<br>p: 2.83E-97 | -- | -- | -- | -- | Calibration=**beta** & TopK=**3**<br>z: +0.86 (Δ +0.15)<br>p: 3.05E-29 |
| Unif × A2 | **beta**<br>z: +0.50 (Δ +0.13)<br>p: 6.89E-86 | -- | **1.0**<br>z: +0.08 (Δ +0.16)<br>p: 3.59E-02 | -- | **all**<br>z: +0.20 (Δ +0.44)<br>p: 4.47E-09 | Calibration=**identity** & TopK=**all**<br>z: +1.00 (Δ +0.20)<br>p: 4.71E-74 |
| Unif × A3 | **identity**<br>z: +0.55 (Δ +0.21)<br>p: 5.39E-34 | **balacc_x_auc**<br>z: +0.43 (Δ +0.13)<br>p: 1.71E-40 | -- | -- | **3**<br>z: +0.70 (Δ +1.28)<br>p: 3.07E-81 | WeightScheme=**balacc_x_auc** & TopK=**3**<br>z: +1.07 (Δ +0.11)<br>p: 2.41E-23 |
| Agg × A0 | **platt**<br>z: +0.63 (Δ +0.12)<br>p: 5.28E-68 | **balacc**<br>z: +0.37 (Δ +0.27)<br>p: 5.53E-09 | **1.0**<br>z: +0.10 (Δ +0.21)<br>p: 5.94E-03 | Best: 0.8<br>z: +0.30 (Δ +0.60)<br>p: 4.59E-16 | -- | Calibration=**platt** & WeightScheme=**balacc**<br>z: +1.17 (Δ +0.29)<br>p: 1.75E-13 |
| Agg × A1 | **identity**<br>z: +1.04 (Δ +1.20)<br>p: 6.15E-81 | -- | -- | -- | **3**<br>z: +0.64 (Δ +1.17)<br>p: 5.81E-65 | Calibration=**identity** & TopK=**3**<br>z: +2.08 (Δ +1.37)<br>p: 1.33E-61 |
| Agg × A2 | -- | **balacc_x_one_minus_brier**<br>z: +0.34 (Δ +0.19)<br>p: 3.47E-14 | **2.0**<br>z: +0.15 (Δ +0.29)<br>p: 8.81E-05 | -- | **3**<br>z: +0.48 (Δ +0.88)<br>p: 1.60E-34 | Calibration=**identity** & WeightScheme=**balacc_x_one_minus_brier**<br>z: +1.01 (Δ +0.16)<br>p: 9.74E-23 |
| Agg × A3 | **identity**<br>z: +1.52 (Δ +1.65)<br>p: 1.88E-302 | -- | -- | -- | -- | Calibration=**identity** & TopK=**3**<br>z: +1.77 (Δ +0.45)<br>p: 2.57E-86 |

## 4. Discussion

In this study we compared plants exposed to different levels of water stress, and the resulting samples were effectively and significantly differentiated by imposed treatments (see Results 3.1), providing a solid basis for the subsequent analyses.

### 4.1 Comparison with prior MK framework

Before addressing the core questions of feature enrichment and ensemble performance, it is instructive to examine how the present dataset behaves relative to our previous work. To this end, we considered the selected features in the baseline configuration (A0), which is equivalent to the pipeline used by Polilli et al. (2026) except for the inclusion of the new experimental data. The comparison shows a broadly consistent pattern, with most morpho-kinematic signatures being confirmed (4 over 6 observations, Results 3.1), highlighting among others the overall importance of linear trend related features. Although, a notable deviation is the reduced prominence of the central rosette region, previously one of the main sources of motion information, in features selected for Tasks B, G, and F focused on chronic stress identification (see Supplementary Table S6). This deviation suggests that the prominence of central rosette features may be dataset-dependent, potentially influenced by factors such as canopy architecture variability, or differential stress acclimation patterns induced by the altered irrigation suspension protocol. A systematic investigation of these factors represents a valuable direction for future work but exceeds the scope of the current additive evaluation.

### 4.2 Feature-space variation study: sectoring and additive levels (A1–A3)

In line with the above premises, feature-selection statistics provide a useful starting point for interpreting the effects of feature-space expansion (i.e., additive levels A1→A3). As the additive level increases, the selected feature sets' sizes declined under the Agg sectoring configuration but remained stable under Unif. Since the fill-rates of A1 and A2 confirm that the newly introduced variables were actively employed we can provide as a plausible interpretation that (for A1 and A2 levels) sectoring by tissue age enhances coherence among features and mitigates the redundancies that may arise in the isogonal (Unif) representation (Figure 5).

Anticipating a more comprehensive discussion on the matter below, and for the sake of the immediately following points, might be beneficial to remark that the three enrichment levels do not contribute equally to model performance. Additions A1 and A2 provided consistent benefits across both sectoring strategies, confirming the general utility of introducing non-linearity (A1) and soil's water-status contextual information through $\Delta t$ (A2). In contrast, A3 yielded measurable gains only under the Unif configuration (Figure 8).

Considering the first addition level (A1), the introduction of external tertiles' slopes markedly altered the composition of the selected feature sets. Across both sectoring strategies, these new non-linear descriptors either surpassed or matched the number of selected rsq features (the linear fit among the baseline descriptors, which in A0 was the only descriptor providing information about the linear/non-linear behavior of the MK prior flattening), effectively increasing the proportion of features informative about linear/non-linear motion trends and cumulatively reaching peaks of 50.0% of the selected pool (Unif A1; See Results 3.2). This enlarged proportion of rsq plus tertiles' slopes remained dominant also across A2 and A3 levels (Supplementary Table S6). Together with the mentioned gain in BA, this underlines both the relevant contribution of the new flattening strategy and the fundamental utility of acceleration linked information about the MK dynamics during the 6h observation window. In addition to that, the prominence of slope_t1 or t3 over rsq, was task dependent: tasks A and F (focused on well-irrigated plants), despite following the same trend for the whole non-linear sub-selection (rsq + slope_t1 or t3), favored the old rsq engineering. Whereas task G, contrasting chronic stress with all other conditions, exhibited a dominant contribution from these tertiles' slopes, showing the highest share of selection consistently throughout all additive levels and in both sectoring schemes. This pattern could suggest the two new slopes as informative for prolonged stress. Yet prudence is demanded to this matter, as the same dominance does not emerge over other tasks managing chronic stress distinctions (Task B and F).

Moving to A2, the inclusion of Δt and its transformations systematically improved performance across tasks and sectoring approaches. The consistency of these gains, heterogenous across tasks, provides insight into Δt's dual role. In schedule-aligned tasks (C, F), where irrigation frequency naturally separates class pairs, Δt could theoretically act as a simple schedule proxy. However, Task F showed no gain ($\Delta BA \approx -1.4$pp), contradicting this expectation. More strikingly, under Agg the schedule-orthogonal Task H improved +3.5pp (Figure 7). This observation, together with the pervasive retention of at least one Δt-related variable in all selected features sets and task-specific preference for $(\Delta t)^2$ *versus* raw Δt, suggests that Δt carries information irreducible to irrigation-schedule alone. Moreover, the same observations are consistent with the hypothesis that Δt defines the context in which plant reacts to the current water status in soil, so that they put in place physiological reactions that previous literature characterized through differences in yield, quality parameters, and vigor indicators (Xu et al., 2010; Acar, 2020; He et al., 2024). While we are observing through MK features. We therefore interpret Δt as a contextual covariate linking morphokinematic responses with recent temporal patterns of water availability. A detailed breakdown of our empirical argument in favor of this interpretation, including task-specific behaviors, is provided in Appendix A.

The third additive level (A3) further extended the feature space by explicitly encoding pairwise interactions between Δt and the other MK, with divergent practical results between Agg and Unif. Specifically, Agg mostly ignored the availability of new predictors and reduced the cardinality of the selection sets resulting in lower performing ensembles, while Unif showed the opposite in all these three aspects (Figure 5 and Figure 8). This behavior indicates that once temporal context (A2) is integrated, explicitly multiplying Δt with other descriptors adds multicollinearity that negatively affects LASSO feature selection method under Agg approach, and suggest two equally plausible interpretations about added information: either A3 adds information that only Unif can leverage trough a larger initial feature set that allows LASSO iterations to retain signaling features, or A3 adds limited information and Unif can leverage the increased feature space trough the slight differences among sectors that encode for canopy asymmetries rather than for biological phenomena. Neither one of these two hypotheses are explored here as this work aims at providing an overview of the gains achievable by introducing the full set of modifications to the original MK analysis pipeline.

Taken together, these results indicate that (i) A1 captures meaningful nonlinear structure already present in the base MK features; (ii) A2 conveys a compact, generalizable context that stabilizes ensemble behaviour; and (iii) A3 adds value only under Unif, possibly due to sector-specific signal concentration. On this basis, we next examine how the choice of ensembling strategy shapes these effects.

### 4.3 Ensembling strategy (HCC vs ALOP)

Given that ALOP ensemble performance depends on its parameter configurations, we conducted a comprehensive evaluation across a wide and diverse array of configurations to ensure an unbiased comparison of its true potential. In our study, we compared the entire performance distribution of ALOP configurations against the scalar metrics of the best-performing HCCs (Figure 8), we observed that in most cases, the entire ALOP distribution surpassed the top HCC. Even in the least favorable scenarios, the ALOP performance median remained higher, confirming that the architecture's advantage is systemic and not the artifact of a single parameter set.

The superior performance of ALOP with respect to HCC classifiers can be attributed to the fundamentally different way these approaches exploit task-specific diversity and manage uncertainty. An HCC leverages structural diversity by arranging binary classifiers into a biologically meaningful hierarchy e.g., identifying stress → distinguishing stress types → discriminating severities (i.e., A→B→C cascade), yet their conditional design makes them inherently prone to error propagation, as any misclassification at higher levels distorts any other subsequent contribution. This limitation has been extensively discussed in the hierarchical classification literature, and while several mitigation strategies have been proposed, none fully prevent the cascading effect of local errors (Silla & Freitas, 2011; Cesa-Bianchi et al., 2006).

By contrast, ALOP performs a parallel fusion of the outputs of all base classifiers, it maintains interpretability by pooling, one class at the time, the opinions of the specialized base models that answer the same biologically valid question as for the HCC; this also eliminates conditional dependencies preventing error propagation (Figure 4). In addition, its adaptive weighting mechanism accounts for differences in empirical reliability across classifiers, effectively down-weighting less consistent models and attenuating large individual errors. In HCC this can be partially addressed by choosing pathways that avoids underperforming tasks (e.g., H→F/C instead of G→A→C as in Polilli et al., 2026). However, that corresponds to a hardcoded on/off switch, that cannot take advantage from more than one perspective (i.e., base model's output) at the time and, therefore, does not contemplate any modulating capacity.

This last advantage is made clearer when deepening the analysis into the ALOPs' performance distribution, from which it emerged that weighted schemes often reached significantly higher BA relative to the unweighted ones (Table 4), this resonate with the principle that predictions should be weighted by their reliability, a notion discussed in opinion pooling philosophical theory as *truth-conduciveness* (Romeijn, 2024) and that is consistent with previous machine learning literature (Kuncheva & Rodríguez, 2014; Large et al., 2019; Iqball & Wani, 2023). This adaptive approach may be particularly valuable in agricultural phenotyping, where classifier reliability can vary substantially across environmental conditions, cultivar, or growth stages. Although this factor can be difficult to specify *a priori*, it could be captured through empirical investigation on a case-by-case basis.

From the same analysis we can derive another important fact about the base models predicted probabilities calibration: a trend emerges at higher addition levels, where identity (i.e., no calibration) yielded better results in both Unif and Agg approaches, while Platt scaling or beta were preferred at lower levels. This observation is more preeminent in Agg, where ALOP parameter schemes including identity were systematically better in A1 and A3 and were not surpassed by other calibration procedures in A2. As Agg sectoring scheme was designed to align with leaves developmental stages, the observation that uncalibrated base models from Agg are effectively utilized by ALOP suggests these sectors capture physiologically meaningful signal, despite their lower BA compared to Unif.

Table 4 reveals complex interactions between feature space, sectoring approach, and ALOP parameters. For instance, base model pruning (task exclusion, TopK) and performance-based weighting schemes were beneficial in Agg across all addition levels, but only at A3 in Unif. This pattern indicates that optimal parametrization is context-dependent rather than universal.

Notably, Agg ALOPs – at the cost of lower BA – exhibit lower interquartile range (IQR) averaged across parameter configurations, suggesting reduced sensitivity to parameter choices, when compared to Unif. The larger and better performing feature sets selected under Unif, may reflect a sensitivity to canopy asymmetries (i.e., Sector 4 vs 6 and Sector 1 vs 3; See Figure 2 and Figure 3) that are biologically irrelevant. By contrast, Agg's biological aggregation by leaf age is unsensitive to such asymmetries, yielding a more parsimonious feature space where signal is concentrated in developmentally coherent regions. This interpretation suggests that biological coherence trades some discriminative power for generalization potential. This robustness (low IQR), however, indicates also that further gains through hyperparameter optimization alone are unlikely. If Agg's lower BA reflects genuine information limitations rather than suboptimal parameterization, future improvements would need to address the root causes: either enriching the training set (to better populate the Agg feature space), expanding the pool of base model architectures (to capture complementary decision boundaries), or refining the sectoring scheme itself (to balance biological interpretability with discriminative power).

As a final remark, for applied phenotyping scenarios, we would recommend the use of ALOP in the Agg-A2 configuration for its balance between accuracy, interpretability, and parsimony. This configuration consistently outperformed HCC architectures while maintaining robustness across LOSO folds, providing the best premises for generalization to unseen replicates within the experimental conditions tested.

### 4.4 Limitations and future work

Despite several advances, this study has limitations. Each methodological component explored here opens new questions. The current sectoring schemes may impose an upper bound on balanced accuracy. The observed importance of non-linear descriptors underscores the need for richer temporal modelling: tertile slopes provide only a sparse approximation of MK dynamics, whereas full-trajectory analysis (all 25 time-points) could unlock deeper understanding. Similarly, Δt appears to carry both physiological and contextual information; future work should quantify these components (e.g., through grouped SHAP analyses).

At the ensemble level, ALOP employed a single base classifier per task, yet in principle each could itself be an ensemble, enabling multi-layer pooling. Moreover, the current ECOC-driven pooling rule does not explicitly account for unequal class coverage across tasks or for composite macro-class sizes, which in more complex designs may affect class-wise comparability and probability calibration. In addition, our evaluation relied on out-of-fold validation within a double-loop scheme; while appropriate for internal validation, embedding this structure within an additional external validation loop (ideally supported by a larger and more heterogeneous dataset) would enable more reliable ALOP parameter optimization. Finally, the study shares some of the general constraints of the previous proof-of-concept: a single species, a controlled environment, and a late growth stage. In the future, broader studies across species, developmental stages, and environmental conditions will be essential to assess the generality of these findings.

Overall, this work extends morpho-kinematic analysis toward a more comprehensive and interpretable framework for crop water-status assessment. It demonstrates that meaningful physiological information can be extracted from time-lapse RGB imaging, while at the same time opening more questions than it resolves, a hallmark of progress in experimental research.

## 5. Conclusion

This work advances morpho-kinematic (MK) analysis toward a more interpretable and generalizable framework for plant water-status assessment. Building upon time-lapse RGB imaging, we demonstrated that methodological refinements at both feature and ensemble levels yield consistent improvements in accuracy and stability. Aggregating canopy sectors by developmental stage (Agg) increased biological coherence, while non-linear temporal descriptors and irrigation-context variables ($\Delta t$) captured dynamic and contextual information neglected by previous baseline.

The introduction of Adaptive Linear Opinion Pooling (ALOP) improved multi-class prediction relative to hierarchical cascades (HCC), eliminating error propagation and leveraging complementary decision perspectives through adaptive weighting. These gains were not limited to a single configuration but were systemic across the parameter space, underscoring the robustness of probabilistic pooling for heterogeneous classifiers.

$\Delta t$ emerged as a distinctive contextual feature with strong evidence for it carrying both schedule-aligned and physiological components, suggesting that morpho-kinematic responses encode the temporal memory of water availability. The Agg and ALOP combination with the inclusion of non-linear temporal descriptors and $\Delta t$ balanced discriminative power, interpretability, and parsimony, offering a reliable architecture for future implementations.

Despite these advances, limitations remain: the study focused on a single species and growth stage under controlled conditions. Broader validation and the adoption of richer temporal models could further strengthen generalization. Overall, this study demonstrates that low-cost RGB imaging, coupled with principled feature engineering and ensemble design, can deliver physiologically meaningful indicators of plant water status, opening avenues for accessible, scalable phenotyping tools in precision agriculture.

**Supplementary Materials**

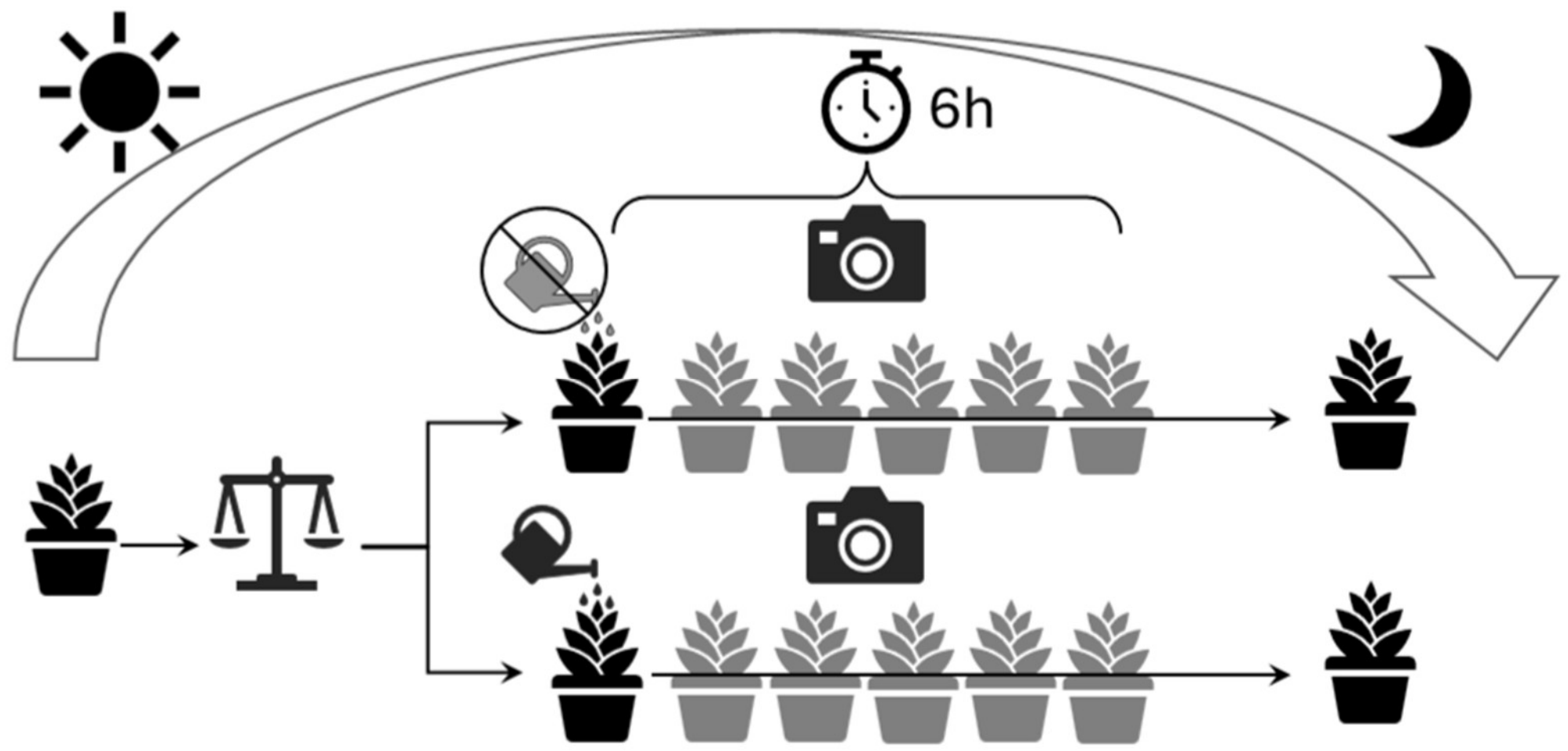

**Supplementary Figure S1:** Observation window and acquisition schedule. For each sample and day, an observation window of 6 hours begins at the daily pot-weighing event (which may or may not include irrigation). Within this window, 25 RGB JPEG images are acquired at 15-minute intervals.

**Supplementary Video S2:** The Supplementary Video S2 follows one sample per experiment, in all its timesteps, for a single day (6h observation window). In the video the panels A1, B1, and C1 decompose the daily ROI definition phase, and they are depicted sequentially for clarity: A1 shows the raw picture at time $n$, in B1 – for visualization purposes only – the canopy is highlighted using a modified $I_{\text{bin}}(x, y, n)$ where $I_{\text{bin}} \in \{0.25,1\}$ and a red dot indicates $\mathbf{c}(n)$, while a yellow circle shows the most distant canopy pixel at time n; panel C1 dynamically updates $\bar{\mathbf{c}}_{k,d}$ (blue dot with a fading trail) and the ROI (area delimited by the green circle), showing the process as it would appear if iteratively recalculated at every timestep $n$.

In the second phase the video shows the real process where the optical flow (panel B2) calculated at timestep $t$ (being t $\in \{1, \dots ,24\}$, the interval between $n = t$ and n = $t + 1$), depicted with magnified vectors, is calculated on the daily (and stable) ROI with no canopy mask applied (Panel A2).

Below we reported two representative frames of Supplementary Video S2 showing the 4th frame (of 25) from a single timelapse of two random samples in phase 1 (ROI definition) and phase 2 (Optical flow calculation).

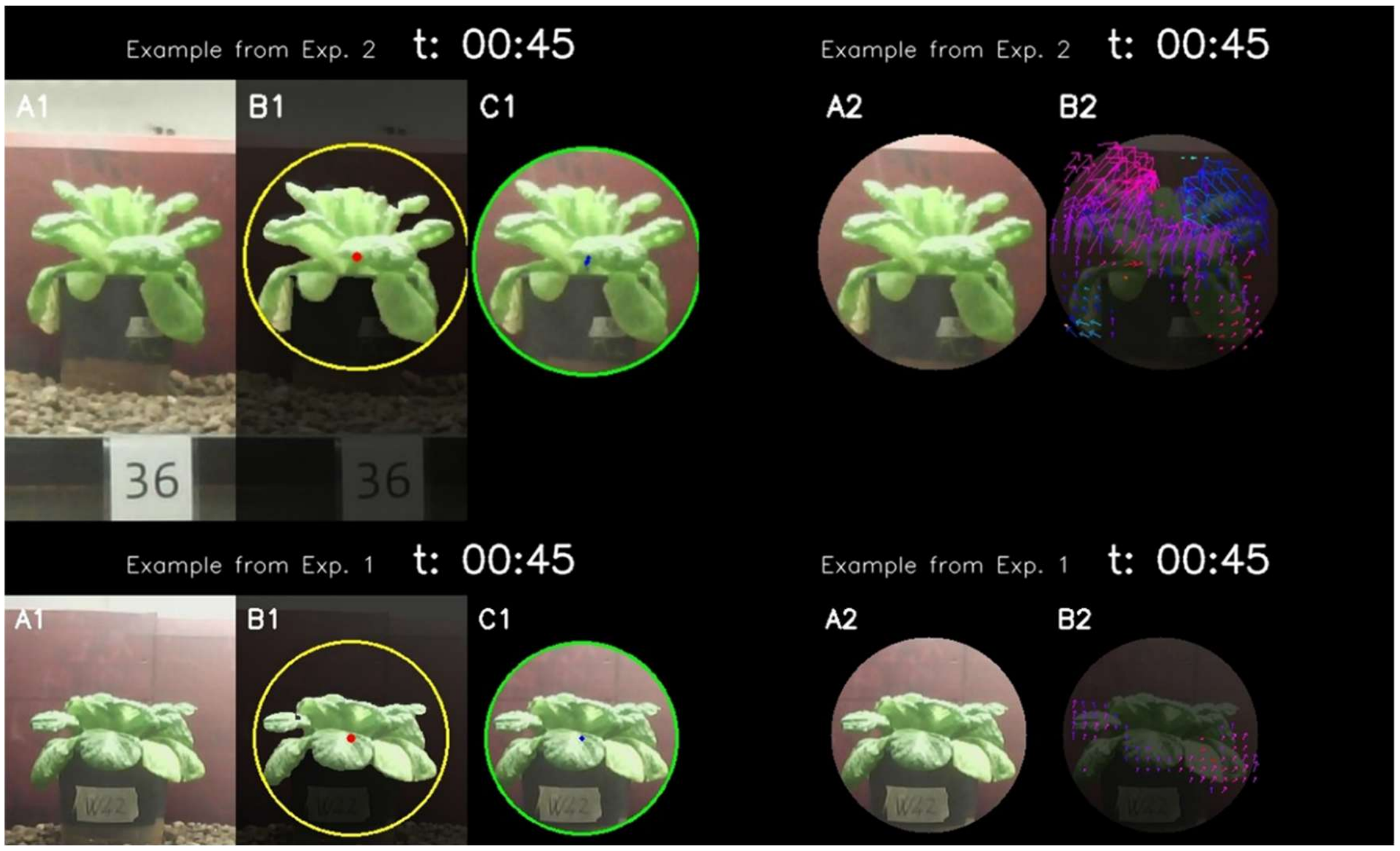

**Supplementary Figure S3:** Two frames of Supplementary Video S2.

**Supplementary Table S4:** Formula for Morpho-Kinematic Features calculation are reported. Let $M_{k,d,s}(x,y) \in \{0,1\}$be the sector mask for sector $s$ within the daily ROI. For each time step $t$ (being $t \in \{1,\dots,24\}$, the interval between frame $n = t$ and frame $n = t + 1$), the flow field is calculated as follows: $\mathrm{F}(x,y,t) = (u(x,y,t), v(x,y,t))$ - where $u$ and $v$ are horizontal and vertical displacements. The flow magnitude is: $Mag(x,y,t) = \sqrt{u(x,y,t)^2 + v(x,y,t)^2}$. Let $\gamma$ be the movement threshold ratio (default $\gamma = 0.01$). Moving pixels are defined within each sector mask $M_{k,d,s}(x,y)$ and for each flow timestep $t$. Let $\gamma$ be the movement-threshold ratio. The moving-pixel set for sample k, day d, sector s, step t is: $P^{mov}_{k,d,s}(t) = \{(x,y): M_{k,d,s}(x,y) = 1 \wedge Mag(x,y,t) > \gamma Rk,d\}$. And its cardinality is: $N^{mov}_{k,d,s}(t) = \mid P^{mov}_{k,d,s}(t) \mid$

| Extended Name | Abbreviation | Formula |
|---|---|---|
| Canopy Density[†] | D | $D^{frame}_{k,d,s}(n) = \frac{\sum_{x,y} I_{\text{bin}}(x,y,n)\, M_{k,d,s}(x,y)}{\sum_{x,y} M_{k,d,s}(x,y)}$<br>$D_{k,d,s}(t) = D^{frame}_{k,d,s}(n = t + 1)$ |
| Movement Threshold | Mth | $Mth_{k,d,s}(t) = \frac{N^{mov}_{k,d,s}(t)}{\sum_{x,y} M_{k,d,s}(x,y)}$ |
| Mean Vertical Movement | VM | $VM_{k,d,s}(t) = \frac{1}{R_{k,d}\, N^{mov}_{k,d,s}(t)} \sum_{(x,y)\in P^{mov}_{k,d,s}(t)} v(x,y,t)$ |
| Mean Horizontal Movement | HM | $HM_{k,d,s}(t) = \frac{1}{R_{k,d}\, N^{mov}_{k,d,s}(t)} \sum_{(x,y)\in P^{mov}_{k,d,s}(t)} u(x,y,t)$ |
| Mean Movement Magnitude | MMag | $MMag_{k,d,s}(t) = \frac{1}{R_{k,d}\, N^{mov}_{k,d,s}(t)} \sum_{(x,y)\in P^{mov}_{k,d,s}(t)} Mag(x,y,t)$ |
| Movement Magnitude Coefficient of Variation[+] | MagCV | $MagCV_{k,d,s}(t) = \frac{\sigma_{Mag}(t)}{\mu_{Mag}(t)}$ |

[†] Canopy density is first defined at the frame level, for sample $k$, day $d$, sector $s$, and frame $n$ (formula above) and then aligned to flow timestep index $t$ as the other MK features (formula below).
[+] Let $\mu_{Mag}(t)$and $\sigma_{Mag}(t)$be mean and standard deviation of $Mag(x,y,t)$over $P^{mov}_{k,d,s}(t)$.

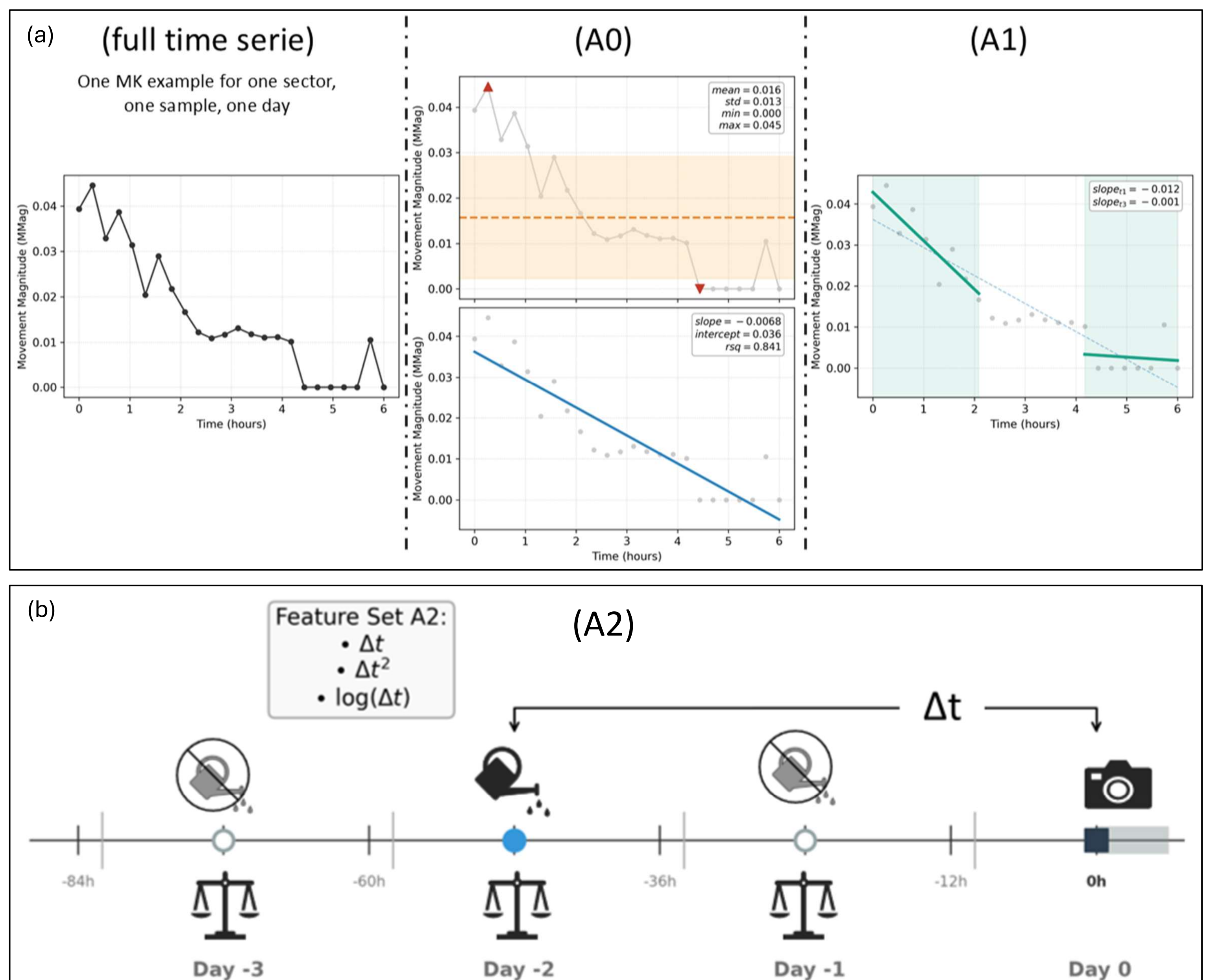

**Supplementary Figure S5**: Example of daily time-series flattening. (a) Left panel, a single full MK time series over the 6h observation window; mid panel, A0 descriptors derived from descriptive statistic and global linear regression (i.e., mean, std, min, max, slope, intercept, rsq); right panel, A1 descriptors adds information about non-linear behavior of the time series (i.e., slope_t1 and slope_t3; green lines). (b) Illustration of $\Delta t$ (timing between last previous irrigation and windows start; see the main text) and A2 features: timeline indicates irrigation events, pot weighing, and the measurement interval used to compute $\Delta t$; A2 includes $\Delta t$, $(\Delta t)^2$, and log $(\Delta t)$.

**Supplementary Table S6:** Features selected per sectoring scheme, task, and additive level. Feature names are coded as follows: (i) referring sector (See main text 2.3); (ii) referring original morpho-kinematic feature (See main text 2.2); (iii) referring flattening category, linreg for linear trend or desc for descriptive; (iv) referring flattening descriptor (See main 2.3); (v) where appropriate for corresponding additive level the tertile slopes follow the "linreg_slope" tag with t1 or t3, irrigation context $\Delta t$ (and transformations) are indicated as "Irrigazione_$\Delta t$", and interactions are indicated by a baseline feature followed by "*" symbol and the related $\Delta t$ transformation.

| Sect. scheme | Task | A lvl | Selected features |
|---|---|---|---|
| Agg | A | A0 | SCR_density_linreg_rsq SCR_horizontal_movement_linreg_rsq SCR_vertical_movement_desc_std SOL_cv_desc_max SOL_density_linreg_intercept SOL_density_linreg_rsq SOL_magnitude_desc_mean SOL_magnitude_linreg_intercept SOL_threshold_desc_max SOL_threshold_linreg_rsq SYL_cv_desc_max SYL_density_linreg_rsq SYL_horizontal_movement_desc_max SYL_horizontal_movement_desc_std SYL_magnitude_desc_min SYL_magnitude_desc_std SYL_threshold_desc_max SYL_vertical_movement_linreg_rsq |
| Agg | B | A0 | SCR_cv_desc_max SCR_cv_linreg_rsq SCR_cv_linreg_slope SCR_density_linreg_slope SCR_horizontal_movement_desc_min SCR_horizontal_movement_linreg_slope SCR_magnitude_linreg_intercept SCR_threshold_desc_max SCR_threshold_desc_min SCR_threshold_linreg_slope SCR_vertical_movement_desc_mean SOL_density_linreg_intercept SOL_density_linreg_rsq SOL_horizontal_movement_desc_max SOL_horizontal_movement_desc_min SOL_horizontal_movement_linreg_intercept SOL_horizontal_movement_linreg_rsq SOL_horizontal_movement_linreg_slope SOL_magnitude_desc_min SOL_threshold_desc_max SOL_threshold_desc_min SOL_vertical_movement_desc_max SOL_vertical_movement_desc_std SOL_vertical_movement_linreg_rsq SYL_cv_desc_mean SYL_cv_desc_min SYL_cv_linreg_rsq SYL_density_linreg_intercept SYL_density_linreg_rsq SYL_density_linreg_slope SYL_horizontal_movement_desc_max SYL_horizontal_movement_linreg_rsq SYL_magnitude_desc_mean SYL_magnitude_desc_min SYL_magnitude_linreg_rsq SYL_threshold_desc_mean SYL_threshold_linreg_rsq |
| Agg | C | A0 | SCR_density_linreg_slope SCR_horizontal_movement_desc_mean SCR_horizontal_movement_linreg_rsq SCR_magnitude_linreg_rsq SCR_threshold_desc_min SCR_threshold_linreg_intercept SCR_vertical_movement_desc_max SCR_vertical_movement_linreg_slope SOL_density_linreg_intercept SOL_horizontal_movement_linreg_rsq SOL_magnitude_desc_min SOL_threshold_desc_min SOL_threshold_linreg_rsq SOL_threshold_linreg_slope SOL_vertical_movement_linreg_rsq SYL_cv_desc_min SYL_density_linreg_intercept SYL_horizontal_movement_linreg_intercept SYL_magnitude_desc_min SYL_threshold_desc_mean SYL_threshold_linreg_slope |
| Agg | F | A0 | SCR_horizontal_movement_linreg_slope SOL_cv_desc_max SOL_cv_desc_std SOL_density_linreg_intercept SOL_density_linreg_rsq SOL_horizontal_movement_linreg_rsq SOL_magnitude_desc_mean SOL_threshold_desc_max SOL_vertical_movement_desc_mean SYL_density_linreg_rsq SYL_horizontal_movement_desc_max SYL_horizontal_movement_desc_mean SYL_horizontal_movement_desc_std SYL_vertical_movement_linreg_rsq |
| Agg | G | A0 | SCR_cv_desc_max SCR_cv_desc_mean SCR_cv_desc_min SCR_density_linreg_rsq SCR_density_linreg_slope SCR_horizontal_movement_desc_max SCR_horizontal_movement_desc_min SCR_horizontal_movement_desc_std SCR_horizontal_movement_linreg_slope SCR_magnitude_desc_max SCR_magnitude_linreg_intercept SCR_threshold_linreg_slope SCR_vertical_movement_desc_max SCR_vertical_movement_desc_mean SCR_vertical_movement_linreg_slope SOL_cv_desc_max SOL_cv_linreg_rsq SOL_density_linreg_intercept SOL_density_linreg_rsq SOL_horizontal_movement_desc_max SOL_horizontal_movement_desc_mean SOL_horizontal_movement_linreg_rsq SOL_horizontal_movement_linreg_slope SOL_magnitude_desc_max SOL_magnitude_desc_mean SOL_magnitude_desc_min SOL_threshold_desc_min SOL_threshold_linreg_slope SOL_vertical_movement_desc_max SOL_vertical_movement_desc_min SOL_vertical_movement_desc_std SOL_vertical_movement_linreg_rsq SYL_cv_desc_mean SYL_cv_linreg_slope SYL_density_linreg_intercept SYL_horizontal_movement_desc_max SYL_horizontal_movement_linreg_intercept SYL_horizontal_movement_linreg_rsq SYL_magnitude_desc_max SYL_magnitude_desc_mean SYL_magnitude_desc_min SYL_threshold_desc_mean SYL_threshold_desc_std SYL_threshold_linreg_intercept SYL_threshold_linreg_rsq SYL_vertical_movement_desc_max SYL_vertical_movement_desc_mean |
| Agg | H | A0 | SCR_cv_desc_max SCR_cv_linreg_rsq SCR_cv_linreg_slope SCR_density_linreg_rsq SCR_density_linreg_slope SCR_horizontal_movement_desc_min SCR_horizontal_movement_linreg_intercept SCR_horizontal_movement_linreg_slope SCR_magnitude_linreg_intercept SCR_threshold_desc_max SCR_threshold_linreg_slope SCR_vertical_movement_desc_mean SOL_density_linreg_intercept SOL_density_linreg_rsq SOL_horizontal_movement_desc_max SOL_horizontal_movement_desc_min SOL_horizontal_movement_linreg_rsq SOL_horizontal_movement_linreg_slope SOL_magnitude_desc_min SOL_threshold_desc_max SOL_threshold_desc_min SOL_vertical_movement_desc_max SOL_vertical_movement_linreg_rsq SYL_cv_desc_mean SYL_density_linreg_intercept SYL_density_linreg_rsq SYL_horizontal_movement_desc_max SYL_horizontal_movement_linreg_rsq SYL_magnitude_desc_mean SYL_magnitude_desc_min SYL_magnitude_linreg_rsq SYL_threshold_desc_mean SYL_threshold_linreg_rsq SYL_vertical_movement_linreg_rsq |
| Agg | A | A1 | SCR_density_linreg_slope_t3 SOL_cv_desc_max SOL_density_linreg_intercept SOL_density_linreg_rsq SOL_density_linreg_slope_t3 SOL_magnitude_desc_mean SOL_magnitude_linreg_intercept SOL_threshold_desc_max SOL_threshold_linreg_rsq SOL_vertical_movement_linreg_slope_t1 SYL_density_linreg_rsq SYL_horizontal_movement_desc_max SYL_horizontal_movement_desc_std SYL_magnitude_desc_min SYL_threshold_desc_max SYL_vertical_movement_linreg_rsq |
| Agg | B | A1 | SCR_cv_desc_max SCR_cv_linreg_rsq SCR_density_linreg_slope SCR_density_linreg_slope_t3 SCR_horizontal_movement_desc_min SCR_horizontal_movement_linreg_slope SCR_horizontal_movement_linreg_slope_t1 SCR_horizontal_movement_linreg_slope_t3 SCR_magnitude_linreg_intercept SCR_magnitude_linreg_slope_t3 SCR_threshold_desc_min SCR_threshold_linreg_slope SCR_vertical_movement_desc_mean SOL_density_linreg_intercept SOL_density_linreg_rsq SOL_horizontal_movement_desc_max SOL_horizontal_movement_linreg_intercept SOL_horizontal_movement_linreg_rsq SOL_horizontal_movement_linreg_slope SOL_horizontal_movement_linreg_slope_t3 SOL_magnitude_desc_min SOL_magnitude_linreg_slope_t3 SOL_threshold_desc_min SOL_vertical_movement_linreg_rsq SYL_cv_desc_mean SYL_cv_linreg_slope_t1 SYL_density_linreg_intercept SYL_density_linreg_rsq SYL_horizontal_movement_linreg_rsq SYL_magnitude_desc_mean SYL_magnitude_desc_min SYL_threshold_desc_mean SYL_threshold_linreg_rsq SYL_vertical_movement_linreg_rsq SYL_vertical_movement_linreg_slope_t1 |
| Agg | C | A1 | SCR_horizontal_movement_desc_mean SCR_horizontal_movement_linreg_rsq SCR_magnitude_linreg_slope_t1 SCR_threshold_desc_mean SCR_threshold_linreg_intercept SCR_threshold_linreg_slope_t1 SCR_vertical_movement_desc_max SCR_vertical_movement_linreg_slope SCR_vertical_movement_linreg_slope_t3 SOL_density_linreg_intercept SOL_density_linreg_slope_t3 SOL_horizontal_movement_linreg_slope_t3 SOL_magnitude_desc_min SOL_threshold_linreg_rsq SOL_threshold_linreg_slope SOL_vertical_movement_linreg_rsq SYL_cv_desc_min SYL_density_linreg_intercept SYL_density_linreg_slope_t3 SYL_horizontal_movement_desc_std SYL_horizontal_movement_linreg_intercept SYL_magnitude_desc_min SYL_threshold_desc_mean SYL_threshold_linreg_slope SYL_threshold_linreg_slope_t1 |
| Agg | F | A1 | SCR_horizontal_movement_linreg_slope SCR_vertical_movement_linreg_rsq SOL_cv_desc_max SOL_density_linreg_rsq SOL_density_linreg_slope_t3 SOL_horizontal_movement_linreg_rsq SOL_magnitude_desc_mean SOL_threshold_desc_max SOL_vertical_movement_linreg_slope_t1 SYL_cv_linreg_slope_t1 SYL_density_linreg_rsq SYL_horizontal_movement_desc_max SYL_horizontal_movement_desc_std SYL_threshold_desc_max |
| Agg | G | A1 | SCR_cv_desc_max SCR_cv_desc_mean SCR_cv_desc_min SCR_cv_linreg_rsq SCR_cv_linreg_slope SCR_density_linreg_slope SCR_density_linreg_slope_t1 SCR_density_linreg_slope_t3 SCR_horizontal_movement_desc_max SCR_horizontal_movement_desc_min SCR_horizontal_movement_linreg_slope SCR_horizontal_movement_linreg_slope_t3 SCR_magnitude_desc_max SCR_magnitude_linreg_intercept SCR_magnitude_linreg_slope_t3 SCR_threshold_linreg_slope SCR_vertical_movement_desc_mean SCR_vertical_movement_linreg_slope SOL_cv_desc_max SOL_density_linreg_rsq SOL_density_linreg_slope_t1 SOL_density_linreg_slope_t3 SOL_horizontal_movement_desc_max SOL_horizontal_movement_linreg_rsq SOL_horizontal_movement_linreg_slope SOL_horizontal_movement_linreg_slope_t3 SOL_magnitude_desc_mean SOL_magnitude_desc_min SOL_magnitude_linreg_slope_t3 SOL_threshold_linreg_slope SOL_vertical_movement_desc_max SOL_vertical_movement_linreg_rsq SOL_vertical_movement_linreg_slope_t1 SYL_cv_desc_mean SYL_cv_linreg_rsq SYL_cv_linreg_slope_t1 SYL_cv_linreg_slope_t3 SYL_density_linreg_intercept SYL_density_linreg_slope_t3 SYL_horizontal_movement_linreg_rsq |

| Sect. scheme | Task | A lvl | Selected features |
|---|---|---|---|
| | | | SYL_horizontal_movement_linreg_slope_t1 SYL_horizontal_movement_linreg_slope_t3 SYL_magnitude_desc_max SYL_magnitude_desc_mean SYL_threshold_desc_mean SYL_threshold_desc_std SYL_threshold_linreg_rsq SYL_vertical_movement_desc_max SYL_vertical_movement_desc_mean SYL_vertical_movement_linreg_slope_t1 SYL_vertical_movement_linreg_slope_t3 |
| Agg | H | A1 | SCR_cv_desc_max SCR_cv_linreg_rsq SCR_cv_linreg_slope SCR_density_linreg_rsq SCR_density_linreg_slope SCR_density_linreg_slope_t3 SCR_horizontal_movement_desc_min SCR_horizontal_movement_linreg_intercept SCR_horizontal_movement_linreg_slope SCR_horizontal_movement_linreg_slope_t1 SCR_horizontal_movement_linreg_slope_t3 SCR_magnitude_linreg_intercept SCR_magnitude_linreg_slope_t3 SCR_threshold_desc_max SCR_threshold_linreg_slope SCR_vertical_movement_desc_mean SCR_vertical_movement_linreg_slope_t1 SOL_cv_linreg_slope_t1 SOL_cv_linreg_slope_t3 SOL_density_linreg_intercept SOL_density_linreg_rsq SOL_horizontal_movement_desc_max SOL_horizontal_movement_desc_min SOL_horizontal_movement_linreg_rsq SOL_horizontal_movement_linreg_slope SOL_magnitude_desc_min SOL_magnitude_linreg_slope_t3 SOL_threshold_desc_max SOL_threshold_desc_min SOL_vertical_movement_desc_max SOL_vertical_movement_linreg_rsq SYL_cv_desc_mean SYL_density_linreg_intercept SYL_density_linreg_rsq SYL_horizontal_movement_desc_max SYL_horizontal_movement_linreg_intercept SYL_horizontal_movement_linreg_rsq SYL_horizontal_movement_linreg_slope_t1 SYL_magnitude_desc_mean SYL_magnitude_desc_min SYL_threshold_desc_mean SYL_threshold_linreg_rsq SYL_vertical_movement_linreg_rsq SYL_vertical_movement_linreg_slope_t1 |
| Agg | A | A2 | Irrigazione SCR_density_linreg_slope_t3 SOL_cv_desc_max SOL_density_linreg_intercept SOL_density_linreg_rsq SOL_density_linreg_slope_t3 SOL_magnitude_desc_mean SOL_magnitude_linreg_intercept SOL_threshold_desc_max SOL_threshold_linreg_rsq SOL_vertical_movement_linreg_slope_t1 SYL_density_linreg_rsq SYL_horizontal_movement_desc_max SYL_horizontal_movement_desc_std SYL_magnitude_desc_min SYL_threshold_desc_max SYL_vertical_movement_linreg_rsq |
| Agg | B | A2 | Irrigazione_dt2 SCR_cv_desc_max SCR_cv_linreg_rsq SCR_density_linreg_slope_t3 SCR_horizontal_movement_desc_std SCR_horizontal_movement_linreg_slope SCR_horizontal_movement_linreg_slope_t3 SCR_magnitude_linreg_intercept SCR_magnitude_linreg_slope_t3 SCR_threshold_desc_min SCR_threshold_linreg_slope SCR_vertical_movement_desc_mean SOL_cv_linreg_slope_t1 SOL_density_linreg_intercept SOL_density_linreg_rsq SOL_horizontal_movement_desc_max SOL_horizontal_movement_linreg_intercept SOL_horizontal_movement_linreg_rsq SOL_horizontal_movement_linreg_slope SOL_horizontal_movement_linreg_slope_t3 SOL_threshold_desc_max SOL_threshold_desc_min SOL_vertical_movement_linreg_rsq SYL_cv_desc_mean SYL_density_linreg_intercept SYL_density_linreg_rsq SYL_density_linreg_slope SYL_horizontal_movement_linreg_rsq SYL_magnitude_desc_mean SYL_magnitude_desc_min SYL_threshold_desc_mean SYL_threshold_linreg_rsq SYL_vertical_movement_linreg_rsq |
| Agg | C | A2 | Irrigazione_dt2 SCR_magnitude_linreg_rsq SCR_threshold_desc_min SCR_threshold_linreg_rsq SCR_vertical_movement_desc_max SCR_vertical_movement_linreg_slope SOL_density_linreg_intercept SOL_horizontal_movement_linreg_slope_t3 SOL_magnitude_desc_min SOL_threshold_desc_min SOL_threshold_linreg_rsq SOL_threshold_linreg_slope SOL_vertical_movement_linreg_rsq SYL_density_linreg_intercept SYL_density_linreg_slope_t3 SYL_horizontal_movement_desc_std SYL_horizontal_movement_linreg_intercept SYL_magnitude_desc_min SYL_threshold_linreg_slope SYL_threshold_linreg_slope_t1 |
| Agg | F | A2 | Irrigazione_dt2 SCR_vertical_movement_linreg_rsq SOL_cv_desc_max SOL_density_linreg_intercept SOL_density_linreg_rsq SOL_density_linreg_slope_t3 SOL_horizontal_movement_linreg_rsq SOL_magnitude_desc_mean SOL_threshold_desc_max SOL_vertical_movement_linreg_slope_t1 SYL_cv_linreg_slope_t1 SYL_density_linreg_rsq SYL_horizontal_movement_desc_std |
| Agg | G | A2 | Irrigazione_dt2 SCR_cv_desc_max SCR_cv_desc_mean SCR_cv_desc_min SCR_cv_linreg_rsq SCR_cv_linreg_slope SCR_density_linreg_slope SCR_density_linreg_slope_t1 SCR_density_linreg_slope_t3 SCR_horizontal_movement_desc_max SCR_horizontal_movement_desc_min SCR_horizontal_movement_linreg_slope SCR_horizontal_movement_linreg_slope_t3 SCR_magnitude_desc_max SCR_magnitude_linreg_intercept SCR_magnitude_linreg_slope_t3 SCR_threshold_linreg_slope SCR_vertical_movement_desc_mean SCR_vertical_movement_linreg_slope SOL_cv_desc_max SOL_density_linreg_intercept SOL_density_linreg_rsq SOL_density_linreg_slope_t1 SOL_density_linreg_slope_t3 SOL_horizontal_movement_desc_max SOL_horizontal_movement_linreg_rsq SOL_horizontal_movement_linreg_slope SOL_horizontal_movement_linreg_slope_t3 SOL_magnitude_desc_mean SOL_magnitude_desc_min SOL_magnitude_linreg_slope_t3 SOL_threshold_linreg_slope SOL_vertical_movement_desc_max SOL_vertical_movement_linreg_rsq SOL_vertical_movement_linreg_slope_t1 SOL_vertical_movement_linreg_slope_t3 SYL_cv_desc_mean SYL_cv_linreg_slope_t1 SYL_cv_linreg_slope_t3 SYL_density_linreg_intercept SYL_horizontal_movement_linreg_rsq SYL_horizontal_movement_linreg_slope_t1 SYL_horizontal_movement_linreg_slope_t3 SYL_magnitude_desc_mean SYL_threshold_desc_mean SYL_threshold_desc_std SYL_threshold_linreg_rsq SYL_vertical_movement_desc_mean SYL_vertical_movement_linreg_slope_t1 |
| Agg | H | A2 | Irrigazione_dt2 SCR_cv_desc_max SCR_cv_linreg_rsq SCR_cv_linreg_slope SCR_density_linreg_rsq SCR_density_linreg_slope SCR_density_linreg_slope_t3 SCR_horizontal_movement_desc_min SCR_horizontal_movement_linreg_intercept SCR_horizontal_movement_linreg_slope SCR_horizontal_movement_linreg_slope_t3 SCR_magnitude_linreg_intercept SCR_magnitude_linreg_slope_t3 SCR_threshold_linreg_intercept SCR_threshold_linreg_slope SCR_vertical_movement_desc_mean SOL_cv_linreg_slope_t1 SOL_cv_linreg_slope_t3 SOL_density_linreg_intercept SOL_density_linreg_rsq SOL_horizontal_movement_linreg_rsq SOL_horizontal_movement_linreg_slope SOL_horizontal_movement_linreg_slope_t1 SOL_threshold_desc_max SOL_threshold_desc_min SOL_vertical_movement_desc_max SOL_vertical_movement_linreg_rsq SYL_cv_desc_mean SYL_density_linreg_intercept SYL_density_linreg_rsq SYL_density_linreg_slope SYL_horizontal_movement_desc_max SYL_horizontal_movement_linreg_rsq SYL_horizontal_movement_linreg_slope_t1 SYL_magnitude_desc_mean SYL_magnitude_desc_min SYL_threshold_desc_mean SYL_threshold_linreg_rsq SYL_vertical_movement_linreg_rsq |
| Agg | A | A3 | Irrigazione SCR_density_linreg_slope_t3 SOL_density_linreg_intercept SOL_density_linreg_rsq SOL_density_linreg_slope_t3 SOL_magnitude_desc_mean SOL_threshold_desc_max SOL_threshold_linreg_rsq SOL_vertical_movement_linreg_slope_t1 SYL_density_linreg_rsq SYL_horizontal_movement_desc_max SYL_horizontal_movement_desc_std SYL_magnitude_desc_min SYL_vertical_movement_linreg_rsq |
| Agg | B | A3 | Irrigazione_dt2 SCR_cv_desc_max SCR_cv_linreg_rsq SCR_density_linreg_slope_t3 SCR_horizontal_movement_linreg_slope SCR_horizontal_movement_linreg_slope_t3 SCR_magnitude_linreg_intercept SCR_magnitude_linreg_slope_t3 SCR_threshold_desc_min SCR_threshold_linreg_slope SCR_vertical_movement_desc_mean SOL_density_linreg_intercept SOL_density_linreg_rsq SOL_horizontal_movement_desc_max SOL_horizontal_movement_linreg_rsq SOL_horizontal_movement_linreg_slope SOL_horizontal_movement_linreg_slope_t1 SOL_horizontal_movement_linreg_slope_t3 SOL_magnitude_linreg_slope_t3 SOL_threshold_desc_min SOL_vertical_movement_linreg_rsq SYL_cv_desc_mean SYL_density_linreg_intercept SYL_density_linreg_rsq SYL_density_linreg_slope SYL_horizontal_movement_linreg_rsq SYL_magnitude_desc_mean SYL_magnitude_desc_min SYL_threshold_desc_mean SYL_threshold_linreg_rsq SYL_vertical_movement_linreg_rsq |
| Agg | C | A3 | Irrigazione_dt2 SCR_horizontal_movement_linreg_rsq SCR_threshold_desc_min SCR_vertical_movement_desc_max SCR_vertical_movement_linreg_slope SOL_density_linreg_intercept SOL_horizontal_movement_linreg_slope_t3 SOL_magnitude_desc_min SOL_threshold_desc_min SOL_threshold_linreg_rsq SOL_threshold_linreg_slope SOL_vertical_movement_linreg_rsq SYL_density_linreg_intercept SYL_density_linreg_slope_t3 SYL_horizontal_movement_desc_std SYL_horizontal_movement_linreg_intercept SYL_magnitude_desc_min SYL_threshold_linreg_slope |
| Agg | F | A3 | Irrigazione_dt2 SOL_cv_desc_max SOL_density_linreg_rsq SOL_density_linreg_slope_t3 SOL_horizontal_movement_linreg_rsq SOL_magnitude_desc_mean SOL_magnitude_linreg_rsq*Irrigazione_dt2 SOL_threshold_desc_max SOL_vertical_movement_linreg_slope_t1 SYL_cv_linreg_slope_t1 SYL_density_linreg_rsq SYL_horizontal_movement_desc_std |
| Agg | G | A3 | Irrigazione_dt2 SCR_cv_desc_max SCR_cv_desc_min SCR_cv_linreg_slope SCR_density_linreg_slope SCR_density_linreg_slope_t3 SCR_horizontal_movement_desc_max SCR_horizontal_movement_linreg_slope SCR_horizontal_movement_linreg_slope_t3 SCR_magnitude_linreg_intercept SCR_magnitude_linreg_slope_t3 SCR_threshold_desc_std SCR_threshold_linreg_slope SCR_vertical_movement_desc_mean SCR_vertical_movement_linreg_slope SOL_cv_desc_max SOL_density_linreg_rsq SOL_density_linreg_slope_t1 SOL_density_linreg_slope_t3 SOL_horizontal_movement_desc_max SOL_horizontal_movement_linreg_rsq SOL_horizontal_movement_linreg_slope SOL_horizontal_movement_linreg_slope_t3 SOL_magnitude_desc_mean SOL_magnitude_desc_min SOL_magnitude_linreg_slope_t3 SOL_vertical_movement_desc_max SOL_vertical_movement_linreg_rsq SOL_vertical_movement_linreg_slope_t1 SYL_cv_desc_mean SYL_cv_linreg_slope_t1 SYL_cv_linreg_slope_t3 SYL_density_linreg_intercept SYL_density_linreg_rsq SYL_horizontal_movement_linreg_intercept SYL_horizontal_movement_linreg_rsq SYL_horizontal_movement_linreg_slope_t1 SYL_horizontal_movement_linreg_slope_t3 SYL_magnitude_desc_mean SYL_magnitude_linreg_rsq SYL_threshold_desc_mean SYL_threshold_desc_std SYL_threshold_linreg_rsq SYL_vertical_movement_desc_mean SYL_vertical_movement_linreg_slope_t1 |
| Agg | H | A3 | Irrigazione_dt2 SCR_cv_desc_max SCR_cv_linreg_rsq SCR_cv_linreg_slope SCR_density_linreg_slope SCR_density_linreg_slope_t3 SCR_horizontal_movement_desc_min SCR_horizontal_movement_linreg_intercept SCR_horizontal_movement_linreg_slope SCR_horizontal_movement_linreg_slope_t3 SCR_magnitude_linreg_intercept SCR_magnitude_linreg_slope_t3 SCR_threshold_desc_min SCR_threshold_linreg_intercept SCR_threshold_linreg_slope SCR_vertical_movement_desc_mean SOL_cv_linreg_slope_t1 SOL_density_linreg_intercept SOL_density_linreg_rsq SOL_horizontal_movement_linreg_rsq SOL_horizontal_movement_linreg_slope SOL_horizontal_movement_linreg_slope_t1 SOL_horizontal_movement_linreg_slope_t3 SOL_magnitude_linreg_slope_t3 SOL_threshold_desc_max SOL_threshold_desc_min SOL_vertical_movement_linreg_rsq SYL_cv_desc_mean SYL_density_linreg_intercept SYL_density_linreg_rsq SYL_horizontal_movement_desc_max SYL_horizontal_movement_linreg_rsq SYL_horizontal_movement_linreg_slope_t1 SYL_magnitude_desc_mean SYL_magnitude_desc_min SYL_threshold_desc_mean SYL_threshold_linreg_rsq SYL_vertical_movement_linreg_rsq |

| Sect. scheme | Task | A lvl | Selected features |
|---|---|---|---|
| Unif | A | A0 | S1_cv_desc_max S1_horizontal_movement_linreg_rsq S1_threshold_desc_max S1_vertical_movement_desc_max S1_vertical_movement_desc_std S2_density_linreg_intercept S2_density_linreg_rsq S3_magnitude_desc_mean S3_threshold_linreg_rsq S4_density_linreg_rsq S4_horizontal_movement_desc_max S4_threshold_desc_max S4_vertical_movement_linreg_rsq S5_horizontal_movement_linreg_rsq S6_cv_desc_mean S6_density_linreg_rsq S6_horizontal_movement_desc_std S6_magnitude_desc_min |
| Unif | B | A0 | S1_density_linreg_rsq S1_horizontal_movement_linreg_rsq S1_horizontal_movement_linreg_slope S2_cv_desc_max S2_cv_desc_mean S2_density_linreg_rsq S2_horizontal_movement_linreg_rsq S2_magnitude_desc_std S2_threshold_desc_mean S2_vertical_movement_linreg_rsq S3_density_linreg_intercept S3_magnitude_desc_min S3_threshold_linreg_slope S3_vertical_movement_desc_max S3_vertical_movement_linreg_rsq S4_density_linreg_intercept S4_horizontal_movement_desc_mean S4_horizontal_movement_linreg_rsq S4_magnitude_desc_mean S4_threshold_linreg_intercept S4_vertical_movement_linreg_slope S5_cv_desc_max S5_density_linreg_rsq S5_density_linreg_slope S5_horizontal_movement_desc_min S5_horizontal_movement_linreg_slope S5_threshold_desc_max S5_threshold_linreg_slope S5_vertical_movement_linreg_slope S6_cv_desc_mean S6_cv_desc_min S6_cv_linreg_rsq S6_density_linreg_intercept S6_density_linreg_slope S6_horizontal_movement_desc_max S6_horizontal_movement_linreg_slope S6_threshold_linreg_rsq S6_vertical_movement_desc_max S6_vertical_movement_linreg_rsq |
| Unif | C | A0 | S1_magnitude_desc_min S2_density_linreg_intercept S2_horizontal_movement_linreg_rsq S2_magnitude_linreg_rsq S3_density_linreg_intercept S3_horizontal_movement_linreg_slope S3_threshold_linreg_slope S3_vertical_movement_linreg_rsq S4_cv_linreg_slope S4_density_linreg_intercept S4_horizontal_movement_linreg_intercept S4_horizontal_movement_linreg_rsq S4_horizontal_movement_linreg_slope S4_magnitude_desc_min S4_threshold_desc_max S4_vertical_movement_desc_max S5_horizontal_movement_linreg_rsq S5_magnitude_linreg_rsq S5_threshold_linreg_intercept S5_vertical_movement_linreg_slope S6_cv_desc_min S6_threshold_linreg_slope |
| Unif | F | A0 | S1_cv_desc_max S1_density_linreg_rsq S1_horizontal_movement_linreg_rsq S2_density_linreg_intercept S2_density_linreg_rsq S3_density_linreg_rsq S3_horizontal_movement_desc_max S3_magnitude_desc_mean S3_threshold_linreg_rsq S4_density_linreg_rsq S5_horizontal_movement_desc_std S6_horizontal_movement_desc_min S6_vertical_movement_linreg_rsq |
| Unif | G | A0 | S1_cv_desc_max S1_density_linreg_rsq S1_horizontal_movement_desc_max S1_horizontal_movement_desc_std S1_horizontal_movement_linreg_rsq S1_horizontal_movement_linreg_slope S1_magnitude_desc_min S1_magnitude_desc_std S2_cv_desc_max S2_density_linreg_rsq S2_horizontal_movement_desc_max S2_horizontal_movement_linreg_rsq S2_magnitude_desc_std S2_threshold_desc_mean S2_vertical_movement_desc_max S2_vertical_movement_desc_min S2_vertical_movement_linreg_rsq S3_density_linreg_intercept S3_horizontal_movement_linreg_rsq S3_magnitude_desc_mean S3_magnitude_desc_min S3_threshold_desc_min S3_threshold_linreg_rsq S3_threshold_linreg_slope S3_vertical_movement_desc_max S3_vertical_movement_linreg_rsq S4_density_linreg_intercept S4_density_linreg_rsq S4_horizontal_movement_desc_mean S4_horizontal_movement_linreg_rsq S4_magnitude_desc_mean S4_magnitude_desc_std S4_threshold_linreg_intercept S4_vertical_movement_desc_mean S4_vertical_movement_desc_min S4_vertical_movement_linreg_slope S5_density_linreg_slope S5_horizontal_movement_desc_std S5_horizontal_movement_linreg_slope S5_magnitude_desc_max S5_threshold_linreg_slope S5_vertical_movement_linreg_slope S6_cv_desc_min S6_cv_linreg_rsq S6_density_linreg_intercept S6_density_linreg_rsq S6_horizontal_movement_desc_max S6_horizontal_movement_linreg_slope S6_magnitude_linreg_rsq |
| Unif | H | A0 | S1_density_linreg_rsq S1_horizontal_movement_linreg_slope S1_magnitude_desc_mean S1_threshold_linreg_rsq S2_density_linreg_intercept S2_density_linreg_rsq S2_horizontal_movement_linreg_rsq S2_threshold_desc_mean S2_vertical_movement_desc_max S2_vertical_movement_linreg_rsq S3_cv_desc_mean S3_density_linreg_slope S3_horizontal_movement_linreg_slope S3_magnitude_desc_min S3_vertical_movement_desc_max S3_vertical_movement_desc_std S3_vertical_movement_linreg_rsq S4_density_linreg_intercept S4_density_linreg_rsq S4_horizontal_movement_linreg_rsq S4_magnitude_desc_mean S4_threshold_desc_mean S4_threshold_linreg_intercept S5_cv_desc_max S5_density_linreg_rsq S5_density_linreg_slope S5_horizontal_movement_desc_min S5_horizontal_movement_linreg_intercept S5_horizontal_movement_linreg_slope S5_threshold_desc_max S5_threshold_linreg_intercept S5_threshold_linreg_slope S5_vertical_movement_linreg_slope S6_cv_desc_mean S6_cv_desc_min S6_cv_linreg_rsq S6_density_linreg_intercept S6_density_linreg_slope S6_horizontal_movement_desc_max S6_horizontal_movement_desc_std S6_horizontal_movement_linreg_slope S6_magnitude_desc_min S6_threshold_linreg_rsq S6_threshold_linreg_slope S6_vertical_movement_desc_max S6_vertical_movement_linreg_rsq |
| Unif | A | A1 | S1_cv_desc_max S1_horizontal_movement_linreg_rsq S1_threshold_desc_max S1_threshold_linreg_rsq S1_vertical_movement_desc_std S2_density_linreg_intercept S3_magnitude_desc_mean S4_density_linreg_rsq S4_threshold_linreg_rsq S4_vertical_movement_linreg_rsq S5_density_linreg_slope_t3 S6_cv_desc_mean S6_density_linreg_rsq S6_density_linreg_slope_t3 S6_horizontal_movement_desc_std S6_magnitude_desc_min |
| Unif | B | A1 | S1_density_linreg_rsq S1_horizontal_movement_linreg_slope S1_magnitude_linreg_slope_t3 S2_cv_desc_max S2_density_linreg_rsq S2_horizontal_movement_linreg_rsq S2_horizontal_movement_linreg_slope_t3 S2_threshold_desc_mean S2_vertical_movement_linreg_rsq S3_density_linreg_intercept S3_vertical_movement_linreg_rsq S4_cv_linreg_slope_t1 S4_density_linreg_intercept S4_density_linreg_rsq S4_horizontal_movement_linreg_rsq S4_magnitude_desc_mean S4_threshold_linreg_intercept S4_threshold_linreg_slope_t3 S5_cv_desc_max S5_density_linreg_rsq S5_density_linreg_slope S5_density_linreg_slope_t1 S5_density_linreg_slope_t3 S5_horizontal_movement_desc_min S5_horizontal_movement_linreg_intercept S5_horizontal_movement_linreg_slope S5_horizontal_movement_linreg_slope_t3 S5_magnitude_linreg_slope_t3 S5_threshold_linreg_slope S5_vertical_movement_linreg_slope S6_cv_linreg_rsq S6_density_linreg_intercept S6_horizontal_movement_desc_max S6_horizontal_movement_linreg_slope S6_magnitude_desc_min S6_vertical_movement_linreg_rsq |
| Unif | C | A1 | S1_magnitude_desc_min S1_threshold_linreg_slope S2_cv_linreg_rsq S2_density_linreg_intercept S3_cv_linreg_slope_t3 S3_density_linreg_intercept S3_density_linreg_slope_t3 S3_threshold_linreg_rsq S3_threshold_linreg_slope S3_vertical_movement_linreg_rsq S4_cv_linreg_slope S4_density_linreg_intercept S4_density_linreg_slope_t3 S4_horizontal_movement_linreg_intercept S4_horizontal_movement_linreg_rsq S4_horizontal_movement_linreg_slope S4_threshold_desc_max S4_vertical_movement_desc_max S5_horizontal_movement_linreg_rsq S5_threshold_linreg_intercept S5_threshold_linreg_slope_t1 S5_vertical_movement_linreg_slope S6_cv_linreg_slope_t3 S6_density_linreg_slope_t1 S6_threshold_linreg_slope |
| Unif | F | A1 | S1_cv_desc_max S1_density_linreg_rsq S1_horizontal_movement_linreg_rsq S2_density_linreg_intercept S2_density_linreg_rsq S3_density_linreg_rsq S3_density_linreg_slope_t3 S3_magnitude_desc_mean S4_cv_linreg_slope_t1 S4_density_linreg_rsq S5_horizontal_movement_desc_std S6_density_linreg_rsq S6_density_linreg_slope_t3 S6_horizontal_movement_desc_min |
| Unif | G | A1 | S1_cv_desc_max S1_density_linreg_rsq S1_density_linreg_slope_t1 S1_horizontal_movement_desc_max S1_horizontal_movement_linreg_rsq S1_horizontal_movement_linreg_slope S1_magnitude_desc_max S1_magnitude_linreg_slope_t3 S1_vertical_movement_linreg_slope_t1 S2_cv_desc_max S2_density_linreg_intercept S2_density_linreg_rsq S2_horizontal_movement_linreg_rsq S2_horizontal_movement_linreg_slope_t3 S2_magnitude_desc_max S2_threshold_desc_mean S2_vertical_movement_linreg_rsq S3_density_linreg_slope_t3 S3_horizontal_movement_linreg_rsq S3_horizontal_movement_linreg_slope_t3 S3_magnitude_desc_mean S3_vertical_movement_linreg_rsq S4_cv_linreg_slope_t1 S4_density_linreg_intercept S4_density_linreg_rsq S4_horizontal_movement_linreg_rsq S4_horizontal_movement_linreg_slope_t1 S4_magnitude_desc_mean S4_magnitude_desc_std S4_threshold_linreg_intercept S4_threshold_linreg_slope_t3 S4_vertical_movement_linreg_slope_t1 S5_density_linreg_slope_t1 S5_horizontal_movement_desc_std S5_horizontal_movement_linreg_slope S5_magnitude_desc_max S5_magnitude_desc_std S5_magnitude_linreg_slope_t3 S5_threshold_linreg_slope S6_cv_linreg_slope_t3 S6_density_linreg_intercept S6_density_linreg_rsq S6_density_linreg_slope_t3 S6_horizontal_movement_desc_max S6_horizontal_movement_desc_mean S6_horizontal_movement_linreg_slope S6_horizontal_movement_linreg_slope_t1 S6_horizontal_movement_linreg_slope_t3 S6_magnitude_linreg_rsq S6_magnitude_linreg_slope_t3 |
| Unif | H | A1 | S1_cv_linreg_slope_t1 S1_cv_linreg_slope_t3 S1_density_linreg_rsq S1_horizontal_movement_desc_mean S1_horizontal_movement_linreg_slope S1_magnitude_desc_mean S1_magnitude_linreg_slope_t3 S1_threshold_linreg_rsq S1_vertical_movement_linreg_rsq S2_density_linreg_rsq S2_horizontal_movement_desc_mean S2_horizontal_movement_desc_min S2_horizontal_movement_linreg_slope_t3 S2_threshold_desc_mean S2_vertical_movement_linreg_rsq S3_cv_desc_mean S3_horizontal_movement_linreg_slope S3_magnitude_desc_min S3_vertical_movement_linreg_rsq S4_cv_linreg_slope_t1 S4_density_linreg_intercept S4_density_linreg_rsq S4_horizontal_movement_linreg_rsq S4_magnitude_desc_mean S4_threshold_linreg_intercept S4_threshold_linreg_slope_t3 S5_cv_desc_max S5_density_linreg_rsq S5_density_linreg_slope S5_density_linreg_slope_t1 S5_density_linreg_slope_t3 S5_horizontal_movement_desc_min S5_horizontal_movement_linreg_slope S5_magnitude_linreg_slope_t3 S5_threshold_linreg_intercept S5_threshold_linreg_slope S5_vertical_movement_linreg_slope S6_cv_desc_mean S6_cv_desc_min S6_cv_linreg_rsq S6_density_linreg_intercept S6_density_linreg_slope S6_horizontal_movement_desc_max S6_horizontal_movement_desc_std S6_horizontal_movement_linreg_slope S6_magnitude_desc_min S6_vertical_movement_desc_max S6_vertical_movement_linreg_rsq S6_vertical_movement_linreg_slope_t3 |
| Unif | A | A2 | Irrigazione S1_cv_desc_max S1_horizontal_movement_desc_std S1_horizontal_movement_linreg_rsq S1_threshold_desc_max S1_vertical_movement_desc_std S1_vertical_movement_linreg_slope_t1 S2_density_linreg_intercept S2_density_linreg_rsq S3_density_linreg_rsq S3_density_linreg_slope_t3 S3_horizontal_movement_desc_max S3_magnitude_desc_mean S4_density_linreg_rsq S4_magnitude_desc_std S4_vertical_movement_linreg_rsq S5_density_linreg_slope_t3 S5_threshold_desc_mean S6_cv_desc_mean S6_density_linreg_rsq S6_density_linreg_slope_t3 S6_horizontal_movement_desc_std S6_magnitude_desc_min |
| Unif | B | A2 | Irrigazione_dt2 S1_density_linreg_rsq S1_horizontal_movement_linreg_slope S1_magnitude_linreg_slope_t3 S2_cv_desc_max S2_density_linreg_rsq S2_horizontal_movement_linreg_rsq S2_horizontal_movement_linreg_slope_t3 S2_threshold_desc_mean S2_vertical_movement_linreg_rsq S3_density_linreg_intercept S3_magnitude_desc_min S3_vertical_movement_linreg_rsq S4_cv_linreg_slope_t1 S4_density_linreg_intercept S4_density_linreg_rsq S4_horizontal_movement_linreg_rsq S4_magnitude_desc_mean S4_threshold_linreg_intercept S4_threshold_linreg_slope_t3 S5_cv_desc_max S5_density_linreg_slope_t1 S5_density_linreg_slope_t3 S5_horizontal_movement_desc_min S5_horizontal_movement_linreg_intercept S5_horizontal_movement_linreg_slope |

| Sect. scheme | Task | A lvl | Selected features |
| --- | --- | --- | --- |
| | | | S5_magnitude_linreg_slope_t3 S5_threshold_linreg_slope S5_vertical_movement_linreg_slope S6_cv_linreg_rsq S6_density_linreg_intercept S6_horizontal_movement_desc_max S6_horizontal_movement_linreg_slope S6_magnitude_desc_min S6_vertical_movement_linreg_rsq |
| Unif | C | A2 | Irrigazione_dt2 S1_magnitude_desc_min S2_cv_linreg_rsq S2_density_linreg_intercept S2_magnitude_linreg_rsq S2_vertical_movement_desc_max S3_density_linreg_intercept S3_density_linreg_slope_t3 S3_threshold_linreg_rsq S3_threshold_linreg_slope S3_vertical_movement_linreg_rsq S4_cv_linreg_slope S4_density_linreg_intercept S4_density_linreg_slope_t3 S4_horizontal_movement_linreg_rsq S4_horizontal_movement_linreg_slope S4_vertical_movement_desc_max S5_vertical_movement_desc_max S5_vertical_movement_linreg_slope S6_density_linreg_intercept S6_magnitude_desc_min S6_threshold_linreg_slope |
| Unif | F | A2 | Irrigazione_dt2 S1_cv_desc_max S1_density_linreg_rsq S1_horizontal_movement_linreg_rsq S1_vertical_movement_linreg_rsq S2_density_linreg_intercept S2_density_linreg_rsq S3_density_linreg_rsq S3_density_linreg_slope_t3 S3_magnitude_desc_mean S4_density_linreg_rsq S6_density_linreg_slope_t3 S6_horizontal_movement_desc_min S6_horizontal_movement_desc_std |
| Unif | G | A2 | Irrigazione_dt2 S1_cv_desc_max S1_density_linreg_rsq S1_density_linreg_slope_t1 S1_horizontal_movement_linreg_rsq S1_horizontal_movement_linreg_slope S1_magnitude_desc_max S1_magnitude_linreg_slope_t3 S1_vertical_movement_desc_mean S1_vertical_movement_linreg_slope_t1 S2_cv_desc_max S2_density_linreg_intercept S2_density_linreg_rsq S2_horizontal_movement_linreg_rsq S2_horizontal_movement_linreg_slope_t3 S2_threshold_desc_mean S2_vertical_movement_linreg_rsq S3_density_linreg_slope_t3 S3_horizontal_movement_linreg_rsq S3_horizontal_movement_linreg_slope_t3 S3_magnitude_desc_mean S3_threshold_linreg_slope S3_vertical_movement_linreg_rsq S4_cv_linreg_slope_t1 S4_density_linreg_intercept S4_density_linreg_rsq S4_horizontal_movement_linreg_rsq S4_horizontal_movement_linreg_slope_t1 S4_magnitude_desc_mean S4_magnitude_desc_std S4_threshold_linreg_intercept S4_threshold_linreg_slope_t3 S5_density_linreg_slope_t1 S5_horizontal_movement_desc_std S5_horizontal_movement_linreg_slope S5_magnitude_desc_std S5_magnitude_linreg_slope_t3 S5_threshold_linreg_slope S6_cv_linreg_slope_t3 S6_density_linreg_intercept S6_density_linreg_rsq S6_density_linreg_slope_t3 S6_horizontal_movement_desc_max S6_horizontal_movement_desc_mean S6_horizontal_movement_linreg_slope S6_horizontal_movement_linreg_slope_t1 S6_horizontal_movement_linreg_slope_t3 S6_magnitude_linreg_rsq S6_magnitude_linreg_slope_t3 |
| Unif | H | A2 | Irrigazione_dt2 S1_cv_linreg_slope_t1 S1_cv_linreg_slope_t3 S1_density_linreg_rsq S1_horizontal_movement_linreg_slope S1_magnitude_linreg_slope_t3 S1_threshold_linreg_rsq S2_density_linreg_rsq S2_horizontal_movement_desc_mean S2_horizontal_movement_desc_min S2_horizontal_movement_linreg_rsq S2_horizontal_movement_linreg_slope_t3 S2_threshold_desc_mean S2_vertical_movement_linreg_rsq S3_magnitude_desc_min S3_vertical_movement_linreg_rsq S4_cv_linreg_slope_t1 S4_density_linreg_intercept S4_density_linreg_rsq S4_horizontal_movement_linreg_rsq S4_magnitude_desc_mean S4_threshold_linreg_intercept S4_threshold_linreg_slope_t3 S5_cv_desc_max S5_density_linreg_rsq S5_density_linreg_slope_t3 S5_horizontal_movement_desc_min S5_horizontal_movement_linreg_intercept S5_horizontal_movement_linreg_slope S5_horizontal_movement_linreg_slope_t3 S5_magnitude_linreg_slope_t3 S5_threshold_linreg_intercept S5_threshold_linreg_slope S5_vertical_movement_linreg_slope S6_cv_desc_mean S6_cv_linreg_rsq S6_density_linreg_intercept S6_density_linreg_slope S6_horizontal_movement_desc_max S6_horizontal_movement_linreg_slope S6_magnitude_desc_min S6_vertical_movement_desc_max S6_vertical_movement_linreg_rsq |
| Unif | A | A3 | Irrigazione Irrigazione_dt2 S1_cv_desc_max S1_horizontal_movement_desc_std S1_horizontal_movement_linreg_rsq S1_threshold_desc_max S1_threshold_linreg_rsq S1_vertical_movement_desc_max S1_vertical_movement_desc_std S1_vertical_movement_linreg_slope_t1 S2_density_linreg_intercept S2_density_linreg_rsq S3_density_linreg_rsq S3_density_linreg_slope_t3 S3_horizontal_movement_desc_max S3_magnitude_desc_mean S4_density_linreg_rsq S4_magnitude_desc_std S4_vertical_movement_linreg_rsq S5_density_linreg_slope_t3 S6_cv_desc_mean S6_density_linreg_rsq S6_density_linreg_slope_t3 S6_horizontal_movement_desc_std S6_magnitude_desc_min S6_magnitude_desc_min*Irrigazione_dt2 |
| Unif | B | A3 | Irrigazione_dt2 S1_horizontal_movement_linreg_slope S1_magnitude_desc_min*Irrigazione_dt2 S1_magnitude_linreg_slope_t3 S2_cv_desc_max S2_density_linreg_rsq S2_horizontal_movement_linreg_rsq S2_horizontal_movement_linreg_slope_t3 S2_threshold_desc_mean S2_vertical_movement_linreg_rsq S3_density_linreg_intercept S3_magnitude_desc_min S3_vertical_movement_linreg_rsq S4_cv_linreg_slope_t1 S4_density_linreg_intercept S4_density_linreg_rsq S4_horizontal_movement_linreg_rsq S4_magnitude_desc_mean S4_threshold_linreg_intercept S5_cv_desc_max S5_density_linreg_slope_t3 S5_horizontal_movement_desc_min S5_horizontal_movement_linreg_intercept S5_horizontal_movement_linreg_slope S5_horizontal_movement_linreg_slope_t3 S5_threshold_linreg_slope S5_vertical_movement_linreg_slope S6_cv_linreg_rsq S6_density_linreg_intercept S6_horizontal_movement_desc_max S6_horizontal_movement_linreg_slope S6_horizontal_movement_linreg_slope_t3 S6_magnitude_desc_min S6_magnitude_desc_min*Irrigazione_dt2 S6_vertical_movement_desc_max S6_vertical_movement_linreg_rsq |
| Unif | C | A3 | Irrigazione_dt2 S1_magnitude_desc_min*Irrigazione_dt2 S2_density_linreg_intercept S2_magnitude_linreg_rsq S2_vertical_movement_desc_max S3_density_linreg_intercept S3_density_linreg_slope_t3 S3_magnitude_linreg_slope_t1 S3_threshold_linreg_rsq S3_threshold_linreg_slope S3_vertical_movement_linreg_rsq S4_cv_linreg_slope S4_density_linreg_intercept S4_density_linreg_slope_t3 S4_horizontal_movement_linreg_rsq S4_horizontal_movement_linreg_slope S4_vertical_movement_desc_max S5_vertical_movement_desc_max S5_vertical_movement_linreg_slope S6_density_linreg_intercept S6_magnitude_desc_min S6_threshold_linreg_slope S6_vertical_movement_linreg_slope_t3 |
| Unif | F | A3 | Irrigazione_dt2 S1_cv_desc_max S1_density_linreg_rsq S1_horizontal_movement_linreg_rsq S1_vertical_movement_linreg_rsq S2_density_linreg_intercept S2_density_linreg_rsq S3_density_linreg_rsq S3_density_linreg_slope_t3 S3_magnitude_desc_mean S4_density_linreg_rsq S6_density_linreg_slope_t3 S6_horizontal_movement_desc_min S6_horizontal_movement_desc_std |
| Unif | G | A3 | Irrigazione_dt2 S1_cv_desc_max S1_density_linreg_rsq S1_density_linreg_slope_t1 S1_horizontal_movement_linreg_rsq S1_horizontal_movement_linreg_slope S1_magnitude_desc_max S1_magnitude_desc_min*Irrigazione_dt2 S1_magnitude_linreg_slope_t3 S2_cv_desc_max S2_density_linreg_intercept S2_density_linreg_rsq S2_horizontal_movement_linreg_rsq S2_horizontal_movement_linreg_slope_t3 S2_magnitude_desc_max S2_magnitude_linreg_slope_t3 S2_threshold_desc_mean S2_vertical_movement_linreg_rsq S3_cv_desc_mean S3_density_linreg_slope_t3 S3_horizontal_movement_desc_std S3_horizontal_movement_linreg_rsq S3_horizontal_movement_linreg_slope_t3 S3_magnitude_desc_mean S3_magnitude_desc_min S3_vertical_movement_linreg_rsq S4_cv_linreg_slope_t1 S4_density_linreg_intercept S4_density_linreg_rsq S4_horizontal_movement_linreg_rsq S4_magnitude_desc_max S4_magnitude_desc_mean S4_threshold_linreg_intercept S5_horizontal_movement_desc_std S5_horizontal_movement_linreg_slope S5_magnitude_desc_std S5_threshold_linreg_slope S6_cv_linreg_rsq S6_cv_linreg_slope_t3 S6_density_linreg_intercept S6_density_linreg_rsq S6_density_linreg_slope_t3 S6_horizontal_movement_desc_max S6_horizontal_movement_desc_mean S6_horizontal_movement_linreg_slope S6_horizontal_movement_linreg_slope_t1 S6_horizontal_movement_linreg_slope_t3 S6_magnitude_linreg_rsq S6_magnitude_linreg_slope_t3 S6_vertical_movement_desc_max |
| Unif | H | A3 | Irrigazione_dt2 S1_cv_linreg_slope_t1 S1_cv_linreg_slope_t3 S1_density_linreg_rsq S1_horizontal_movement_linreg_slope S1_magnitude_desc_min*Irrigazione_dt2 S1_magnitude_linreg_slope_t3 S1_threshold_linreg_rsq S2_density_linreg_rsq S2_horizontal_movement_linreg_slope_t1 S2_horizontal_movement_linreg_slope_t3 S2_threshold_desc_mean S2_vertical_movement_linreg_rsq S3_horizontal_movement_linreg_slope S3_magnitude_desc_min S3_vertical_movement_linreg_rsq S4_cv_linreg_slope_t1 S4_density_linreg_intercept S4_density_linreg_rsq S4_horizontal_movement_linreg_rsq S4_magnitude_desc_mean S4_magnitude_linreg_slope_t3 S4_threshold_linreg_intercept S5_cv_desc_max S5_density_linreg_rsq S5_density_linreg_slope_t3 S5_horizontal_movement_desc_min S5_horizontal_movement_linreg_intercept S5_horizontal_movement_linreg_slope S5_horizontal_movement_linreg_slope_t3 S5_magnitude_linreg_slope_t3 S5_threshold_linreg_intercept S5_threshold_linreg_slope S5_vertical_movement_linreg_slope S6_cv_desc_mean S6_cv_linreg_rsq S6_density_linreg_intercept S6_horizontal_movement_desc_max S6_horizontal_movement_linreg_slope S6_magnitude_desc_min S6_magnitude_desc_min*Irrigazione_dt2 S6_vertical_movement_desc_max S6_vertical_movement_linreg_rsq |

**Supplementary Table S7:** Comparison between the two groups of treatments A={FC, SM} and B={SS, SC} under three transformations of normalized Δt: Δt, $(\Delta t)^2$, and log(Δt). Table reports Mann–Whitney U results (MWU) with sample sizes (n_A, n_B), two-sided *p*-value, the Holm-adjusted *p*-value, the rank-biserial effect size (RB_effect); Shapiro–Wilk (SW) and Levene's test *p*-values; and OOF performance of a 1-D logistic regression (5-fold stratified CV): ROC AUC (mean ± sd) and Balanced Accuracy (mean ± sd).

| Feature | MWU | n_A, n_B | *p*-value raw | *p*-value Holm | RB_effect | SW_A *p*-value | SW_B *p*-value | Levene's *p*-value | Balacc (mean ± sd) | ROC AUC (mean ± sd) |
|---|---|---|---|---|---|---|---|---|---|---|
| Δt | 1548 | 72, 72 | 2.9 E-05 | 8.0 E-05 | 0.403 | 4.4 E-06 | 2.2 E-04 | 4.0 E-03 | 0.66 ± 0.08 | 0.69 ± 0.08 |
| $(\Delta t)^2$ | 1548 | 72, 72 | 2.9 E-05 | 8.0 E-05 | 0.403 | 4.6 E-10 | 8.2 E-09 | 5.1 E-05 | 0.60 ± 0.10 | 0.69 ± 0.08 |
| log(Δt) | 1548 | 72, 72 | 2.9 E-05 | 8.0 E-05 | 0.403 | 1.9 E-14 | 4.5 E-15 | 9.3 E-01 | 0.61 ± 0.05 | 0.69 ± 0.08 |

**Appendix A: Evidence for Context-Defining Component of Δt**

Definition: *Δt is the elapsed time – normalized to (0,1) – between the last irrigation preceding the daily pot-weighing; if irrigation occurs that day, Δt is measured up to the check and is not reset.*

Use*: Δt along $(\Delta t)^2$ and log(Δt) were added as available features in the addition level A2*

As noted in the main text, Δt carries a schedule-aligned component due to differential irrigation frequencies ({FC, SM} daily vs. {SS, SC} every other day; Figure 6; Table AA). Here we examine evidence for an additional context-defining component.

If Δt carried only schedule-aligned information, we would expect performance gains (A1→A2) to concentrate in tasks where classes separate along schedule boundaries. To test this, we categorize tasks based on class-schedule synchronization: schedule-aligned (Task C, F: classes separate into {FC, SM} vs. {SS, SC}), partially aligned (Task A, B, G: one class mixes schedules), and orthogonal (Task H: both classes mix schedules). This organization matches by-design the amount of information gainable by the schedule-aligned signal alone when made available by Δt features (in A2), therefore under a pure schedule derived signal, we predict monotonically decreasing gains: aligned (a) > partially (pa) > orthogonal (o).

With this framework in mind, we can highlight the first piece of evidence: feature selection retained a Δt-related feature in all tasks (Supplementary Table S6; Figure 5), including schedule-orthogonal Task H where schedule-based information provides no class separation. This universal selection, robust across LOSO folds, suggests Δt contributes beyond its schedule-aligned component. The second piece of evidence directly contradicts predictions from a schedule-only model. Table AA shows that balanced accuracy (BA) dynamics from A1 to A2 does not follow the predicted monotonic decline. In fact, (i) Task H (schedule-orthogonal) gained +3.5pp in Agg despite schedule providing no class separation; (ii) the highest average gains in Unif occurred in partially aligned tasks (B: +7.2, G: +7.5); (iii) aligned Task F showed no gain. This pattern is inconsistent with Δt acting purely as a schedule proxy.

**Table AA**: Performance gains (ΔBA, percentage points) from A1 to A2 across tasks, categorized by irrigation schedule alignment. A schedule-only signal would predict monotonic decline (a > pa > o).

| Task | Schedule alignment | ΔBA_AGG | ΔBA_Unif |
|---|---|---|---|
| TaskC | a | 7.402597 | 4.545455 |
| TaskF | a | -1.38889 | -1.38889 |
| TaskA | pa | -4.05983 | -1.44231 |
| TaskB | pa | 4.330065 | 7.189542 |
| TaskG | pa | 2.831197 | 7.478632 |
| TaskH | o | 3.513072 | -0.77614 |

The third piece of evidence is less of a statement for a tout-court-role of Δt and more of an indication for how and where that role exerts its effect the most, we argue that the following observation are consistent with the previous and open a window for further deeper studies.

Two tasks showed anomalous A1→A2 behavior warranting closer examination: Task A and Task F, which are the only two tasks having the well irrigated treatment (FC) isolated in one class.

Task F (FC vs. SC): Despite being schedule-aligned, Task F showed no gain (ΔBA ≈ -1.4pp both approaches). This suggests that even where schedule information is available, Δt does not uniformly improve performance when the physiological contrast is primarily about stress vs. well-watered status.

Task A (FC vs. {SC, SS, SM}): Task A exhibited negative ΔBA (-4.1 Agg) despite more helpful (wrong→right: 4.3%) than harmful (right→wrong: 3.6%) switches. This apparent paradox is resolved by examining class-specific changes: ΔTrue Positive Rate (identifying FC) = -13.9pp; ΔTrue Negative Rate (identifying stress) = +5.8pp. Under class imbalance (FC is minority positive class), the TPR loss dominated balanced accuracy despite the TNR gain.

This asymmetry is particularly revealing considering feature selection: Task A uniquely favored raw Δt, whereas all other tasks selected $(\Delta t)^2$. Since Δt is normalized [0,1], squaring emphasizes larger values-specifically, the prolonged dry intervals experienced only by stress treatments (SC, SS). The raw-vs-squared dichotomy suggests that:

1. In tasks distinguishing among stress types (B, C, G, H), high Δt values provide discriminative context about recovery/adaptation state → $(\Delta t)^2$ amplifies this signal
2. In Task A (well-watered vs. any stress), high Δt values are irrelevant for the positive class (FC never experiences them), and moderate Δt values within FC remain physiologically "short" relative to stress-induced perturbations → raw Δt avoids over-emphasizing uninformative high values

Crucially, Task A's class structure (being partially aligned) creates a conflict between task definition and schedule grouping: FC and SM are in opposite task classes but same schedule group {FC, SM}. When Δt carries only schedule information, this creates confusion; when Δt additionally carries stress-context information (where SM's history differs from FC despite shared schedule), it aids stress discrimination (TNR↑) but not FC identification (TPR↓). The observation that Δt is more effective for distinguishing stress types than for separating well-watered from stressed aligns with plant physiological literature showing that drought duration modulates recovery capacity and stress responses non-linearly (Xu et al., 2010; Acar, 2020; He et al., 2024).

Taken together, these three independent lines of evidence - universal feature selection, heterogeneous schedule-violating gains, and task-specific transformation preferences - converge on the interpretation that Δt carries both schedule-aligned and context-defining information. We adopt this as our working hypothesis for interpreting Δt's role in the classification models presented in the main text.

We emphasize that this remains a hypothesis requiring formal validation. Rigorous quantification would require controlled experiments (e.g., group-aware importance measures like group-SHAP, or schedule-preserving perturbations of Δt). Such tests, while beyond the scope of this study, represent a clear priority for future work to disentangle the two components without introducing artifacts.